\documentclass[aps,prd,nofootinbib,showpacs,superscriptaddress, preprint]{revtex4-1}
\pdfoutput=1
\usepackage[utf8]{inputenc}
\usepackage{amsmath,amssymb}
\usepackage{epsfig}
\usepackage{graphicx}
\usepackage[usenames,dvipsnames]{color}
\usepackage[colorlinks,citecolor=blue]{hyperref}
\usepackage{color}
\usepackage{subfigure}

\begin{document}
\title{TeV scale leptogenesis, inflaton dark matter and neutrino mass in a scotogenic model}

\author{Debasish Borah}
\email{dborah@iitg.ac.in}
\affiliation{Department of Physics, Indian Institute of Technology Guwahati, Assam 781039, India}
\author{P. S. Bhupal Dev}
\email{bdev@wustl.edu}
\affiliation{Department of Physics and McDonnell Center for the Space Sciences, Washington University,
St. Louis, Missouri 63130, USA}
\author{Abhass Kumar}
\email{abhasskumar@hri.res.in}
\affiliation{Harish-Chandra Research Institute, Chhatnag Road, Jhunsi, Allahabad 211019, India}
\affiliation{Homi Bhabha National Institute, Training School Complex, Anushakti Nagar, Mumbai 400094, India}

\begin{abstract}
We consider the scotogenic model, where the standard model (SM) is
extended by a scalar doublet and three $Z_2$ odd SM-singlet fermions ($N_i$, $i=1,2,3$), all odd under an
additional $Z_2$ symmetry, as a unifying framework for simultaneous explanation of inflation, dark matter, baryogenesis and neutrino mass. The inert doublet is coupled nonminimally to gravity and forms the inflaton. The lightest neutral particle of this doublet later becomes the dark matter candidate. Baryogenesis is achieved via leptogenesis by the decay of $N_1$ to SM leptons and the inert doublet particles. Neutrino masses are generated at the one-loop level. Explaining all these phenomena together in one model is very economic and gives us a new set of constraints on the model parameters. We calculate the inflationary parameters like spectral index, tensor-to-scalar ratio and scalar power spectrum, and find them to be consistent with the Planck 2018 constraints. We also do the reheating analysis for the inert doublet decays/annihilations to relativistic, SM particles. We find that the observed baryon asymmetry of the Universe can be obtained and the sum of light neutrino mass bound can be satisfied for the lightest $Z_2$ odd singlet fermion of mass around 10 TeV, dark matter in the mass range 1.25--1.60 TeV, and the lepton number violating quartic coupling between the SM Higgs and the inert doublet in the range of $6.5\times10^{-5}$ to $7.2\times 10^{-5}$.
\end{abstract}
\maketitle

\section{Introduction}
There has been significant progress in the last few decades in gathering evidence for the presence of a mysterious, nonluminous, nonbaryonic form of matter, known as dark matter (DM), in the present Universe~\cite{Tanabashi:2018oca}.  From the early galaxy cluster observations~\cite{Zwicky:1933gu}, observations of galaxy rotation curves~\cite{Rubin:1970zza}, the more recent observation of the bullet cluster~\cite{Clowe:2006eq} and the latest cosmological data provided by the Planck satellite~\cite{Aghanim:2018eyx}, it is now very much certain that approximately $27\%$ of the present Universe is composed of DM, which is about five times more than the ordinary luminous or baryonic matter. Certain criteria to be satisfied by particle candidates for DM can be found in \cite{Taoso:2007qk} which rule out all the standard model (SM) particles as DM candidates. Among different beyond standard model (BSM) proposals for DM~\cite{Feng:2010gw}, the weakly interacting massive particle (WIMP) paradigm remains the most widely studied scenario where a DM candidate typically with electroweak (EW) scale mass and interaction rate similar to EW interactions can give rise to the correct DM relic abundance, a remarkable coincidence often referred to as the \textit{WIMP Miracle}~\cite{Kolb:1990vq}.

Apart from DM, the baryonic part of the Universe itself provides another puzzle -- an abundance of baryons over antibaryons. The dynamical production of a remnant baryon asymmetry (excess baryons over antibaryons) requires certain conditions to be fulfilled if the Universe is assumed to have started in a baryon-symmetric manner. These conditions, known as the Sakharov conditions~\cite{Sakharov:1967dj} require baryon number (B) violation, C and CP violation and departure from thermal equilibrium, not all of which can be fulfilled in the required amounts within the SM alone. Out-of-equilibrium decay of a heavy particle leading to the generation of baryon asymmetry of the Universe (BAU) has been a well-known mechanism for baryogenesis~\cite{Weinberg:1979bt, Kolb:1979qa}. One interesting way to implement such a mechanism is leptogenesis~\cite{Fukugita:1986hr}, where a net leptonic asymmetry is generated first which gets converted into baryon asymmetry through $(B+L)$-violating EW sphaleron transitions~\cite{Kuzmin:1985mm}. For the lepton asymmetry to be converted into baryon asymmetry, it is important that the processes giving rise to the leptonic asymmetry freeze out before the onset of the sphaleron transitions to prevent wash-out of the asymmetry~\cite{Fong:2013wr}. An interesting feature of this scenario is that the required lepton asymmetry can be generated through CP violating out-of-equilibrium decays of the same heavy fields that take part in the seesaw mechanism~\cite{Minkowski:1977sc, Mohapatra:1979ia, Yanagida:1979as, GellMann:1980vs, Glashow:1979nm, Schechter:1980gr} that explains the origin of tiny neutrino masses~\cite{Tanabashi:2018oca}, another observed phenomenon the SM fails to address.

The assumption that the Universe should have started in a baryon-symmetric manner is bolstered by inflation, another widely studied problem in cosmology. Originally proposed to solve the horizon, flatness and unwanted relic problem in cosmology \cite{Guth:1980zm, Linde:1981mu}, the inflationary paradigm is also supported by the adiabatic and scale invariant perturbations observed in the cosmic microwave background \cite{Komatsu:2010fb, Akrami:2018odb}. Any baryon asymmetry present in the Universe before inflation would be washed out at the end of inflation due to the exponential expansion of the Universe which would dilute any previous information. Over the years, a variety of inflationary models have been studied with different levels of success~\cite{Mazumdar:2010sa}. Chaotic inflation \cite{Linde:1983gd} models were one of the earliest and simplest that used power law potentials like $m^2\phi^2+\lambda\phi^4$ with a scalar $\phi$. These models were not very accurate at explaining the observations. Another class of models use the Higgs as the inflaton \cite{Bezrukov:2007ep, Bezrukov:2010jz}, the particle responsible for inflation. These models often suffer from problems of vacuum stability \cite{Sher:1988mj} and nonunitarity \cite{Lerner:2009na} as well as being inadequate for combining inflation with other cosmological problems like DM and baryogenesis. A way out is adding an extra stabilizing scalar which acts as the inflaton.

In this work, we consider the possibility of connecting the above three phenomena, namely, DM, baryon asymmetry and inflation, which may seem unrelated to each other, within the framework of a simple, nonsupersymmetric particle physics model that also explains nonzero neutrino masses.\footnote{See Refs.~\cite{Shaposhnikov:2006xi, Allahverdi:2007wt, Kohri:2009ka, Boucenna:2014uma, Salvio:2015cja, Ballesteros:2016euj} for other examples which connect all these phenomena in a single unifying framework. A comparative discussion of the scotogenic scenario considered here vis-\'{a}-vis these alternative constructions is given at the end of Sec.~\ref{sec7}.} The model is based on the scotogenic framework~\cite{Ma:2006km} where the SM is extended by three copies of SM-singlet $Z_2$ odd fermions and an additional Higgs doublet, all of which are odd under an unbroken $Z_2$ symmetry, leaving the possibility of the lightest $Z_2$-odd particle to be a stable DM candidate. The additional Higgs doublet is also often called the inert doublet, as it does not develop a vacuum expectation value (VEV). These $Z_2$-odd particles also take part in generating light neutrino masses at one-loop level. We consider the $Z_2$-odd scalar field, namely the inert Higgs doublet to play the role of DM and inflaton simultaneously while the $Z_2$-odd fermions create the leptonic asymmetry through out-of-equilibrium decay into SM leptons and inert Higgs doublet.\footnote{The possibility of a single field or particle playing the role of inflaton and DM was first pointed out in Refs.~\cite{Kofman:1994rk, Kofman:1997yn} and was taken up for detailed studies in several subsequent works~\cite{Liddle:2006qz, Cardenas:2007xh, Panotopoulos:2007ri, Liddle:2008bm, Bose:2009kc, Lerner:2009xg, Okada:2010jd, DeSantiago:2011qb, Lerner:2011ge, delaMacorra:2012sb, Khoze:2013uia,  Kahlhoefer:2015jma, Bastero-Gil:2015lga, Tenkanen:2016twd, Choubey:2017hsq, Heurtier:2017nwl, Hooper:2018buz,Daido:2017tbr,Daido:2017wwb}.} To keep the scenario minimal and simple, we consider a variant of Higgs inflation \cite{Bezrukov:2007ep, Bezrukov:2010jz} where the inert Higgs doublet field having nonminimal coupling to gravity can serve as the inflaton \cite{Choubey:2017hsq}, can reheat the Universe after inflation giving rise to a radiation dominated phase and also play the role of DM with the correct relic abundance and satisfying other DM related constraints like direct detection. We extend this scenario by introducing three $Z_2$ odd SM-singlet fermions to account for the baryon asymmetry in the Universe. The reheating after inflation produces these fermions in thermal equilibrium which are responsible for generating the lepton asymmetry at a temperature approximately equal to the lightest such fermion mass $T \sim M_1$. We find that the required amount of lepton asymmetry can be produced for $M_1 \sim 10$ TeV within a vanilla leptogenesis framework having hierarchical $Z_2$ odd singlet fermionic masses while satisfying the constraints from light neutrino masses. This also agrees with the recent study of low scale leptogenesis in scotogenic model \cite{Hugle:2018qbw}. We obtain values for inflationary parameters like the spectral index $n_s=0.9678$, the tensor-to-scalar ratio $r=0.0029$ which are consistent with the 2018 Planck constraints~\cite{Akrami:2018odb}. We use the scalar power spectrum to obtain the relation between the quartic self-coupling $\lambda_2$ of the inert doublet and its coupling to gravity. Reheating analysis allows us to get a lower bound on $\lambda_2\gtrsim \frac{1}{60}$. Successful baryogenesis via leptogenesis and DM relic abundance is obtained for the DM mass range of 1.25--1.60 TeV and the lepton-number-violating quartic coupling $\lambda_5$ between the SM Higgs and the inert doublet in the range of $6.5\times10^{-5}$ to $7.2\times 10^{-5}$, which also satisfies the sum of neutrino mass bounds from Planck. 

The rest of this paper is organized as follows. In Sec. \ref{sec1}, we briefly summarize the minimal scotogenic model. In Sec.~\ref{sec:nu}, we review how the neutrino mass is generated in this model. This is followed by discussions on inflation and reheating in Secs.  \ref{sec2} and \ref{sec3}, respectively. We discuss the details of DM and baryogenesis through leptogenesis in Secs. \ref{sec4} and \ref{sec5}, respectively. We then summarize the study of renormalization group (RG) evolution of model parameters in section \ref{sec6} and then conclude in section \ref{sec7}.

\section{The Scotogenic Model}
\label{sec1}
As mentioned earlier, the minimal scotogenic model is the extension of the SM by three copies of SM-singlet $Z_2$ odd fermions $N_i$ (with $i=1,2,3$) and one $SU(2)_L$-doublet scalar field $\Phi_2$ (called inert doublet), all being odd under a $Z_2$ symmetry, while the SM fields remain $Z_2$-even, i.e. under the $Z_2$-symmetry, we have  
\begin{align}
N \rightarrow -N, \quad  \Phi_2 \rightarrow -\Phi_2, \quad \Phi_{1} \rightarrow \Phi_{1}, \quad \Psi_{\rm SM} \to 
\Psi_{\rm SM} \, ,
\label{eq:Z2}
\end{align}
where $\Phi_1$ is the SM Higgs doublet and $\Psi_{\rm SM}$'s stand for the SM fermions. This $Z_2$ symmetry, though \textit{ad hoc}  in this minimal setup, could be realized naturally as a subgroup of a continuous gauge symmetry like $U(1)_{B-L}$ with nonminimal field content \cite{Dasgupta:2014hha,Das:2017ski}. The $Z_2$ symmetry and the corresponding charges of the fields prevent the SM fermion couplings with the additional scalar $\Phi_2$ at renormalizable level, thus making it inert, as the name ``inert doublet" suggests. However, the SM leptons can couple to the inert doublet $\Phi_2$ via the $Z_2$ odd SM-singlet fermions. The $Z_2$-symmetry still prevents the neutral component of $\Phi_2$ from acquiring a nonzero VEV. This eventually forbids the generation of light neutrino masses at tree level through the conventional type-I seesaw mechanism \cite{Minkowski:1977sc, Mohapatra:1979ia, Yanagida:1979as, GellMann:1980vs, Glashow:1979nm, Schechter:1980gr}.

The scalar sector of the model is same as the inert Higgs doublet model (IHDM) \cite{Deshpande:1977rw}, a minimal extension of the SM in order to accommodate a DM candidate~\cite{Ma:2006km, Dasgupta:2014hha, Cirelli:2005uq, Barbieri:2006dq, Ma:2006fn,  LopezHonorez:2006gr,  Hambye:2009pw, Dolle:2009fn, Honorez:2010re, LopezHonorez:2010tb, Gustafsson:2012aj, Goudelis:2013uca, Arhrib:2013ela, Diaz:2015pyv, Ahriche:2017iar}. Due to the $Z_2$ symmetry given by Eq.~\eqref{eq:Z2}, this prevents linear and trilinear couplings of $\Phi_2$ with the SM Higgs. Therefore, if the bare mass-squared term of $\Phi_2$ is chosen positive definite, its neutral components do not acquire any VEV even after electroweak symmetry breaking (EWSB). This ensures the stability of the lightest component of $\Phi_2$, irrespective of its mass, on cosmological scale. If this lightest component is electromagnetically neutral, then this can be a good DM candidate, if other relevant constraints are satisfied. The scalar potential of the model involving the SM Higgs doublet $\Phi_1$ and the inert doublet $\Phi_2$ can be written as
\begin{align}
V(\Phi_1,\Phi_2) & \ = \   \mu_1^2|\Phi_1|^2 +\mu_2^2|\Phi_2|^2+\frac{\lambda_1}{2}|\Phi_1|^4+\frac{\lambda_2}{2}|\Phi_2|^4+\lambda_3|\Phi_1|^2|\Phi_2|^2 \nonumber \\
& \qquad +\lambda_4|\Phi_1^\dag \Phi_2|^2 + \left[\frac{\lambda_5}{2}(\Phi_1^\dag \Phi_2)^2 + \text{H.c.}\right] \, . \label {c}
\end{align}
To ensure that none of the neutral components of the inert Higgs doublet acquire a nonzero VEV, $\mu_2^2 >0$ is assumed. This also prevents the $Z_2$ symmetry from being spontaneously broken. The EWSB occurs due to the nonzero VEV acquired by the neutral component of $\Phi_1$. After the EWSB, these two scalar doublets can be written in the following form in the unitary gauge:
\begin{equation}
\Phi_1 \ = \ \begin{pmatrix} 0 \\  \frac{ v +h }{\sqrt 2} \end{pmatrix} , \qquad \Phi_2 \ = \ \begin{pmatrix} H^\pm\\  \frac{H^0+iA^0}{\sqrt 2} \end{pmatrix} \, ,
\label{eq:idm}
\end{equation}
where $h$ is the SM-like Higgs boson, $H^0$ and $A^0$ are the CP-even and CP-odd scalars, and $H^\pm$ are the charged scalars from the inert doublet. The masses of the physical scalars at tree level can be written as
\begin{eqnarray}
m_h^2 & \ = \ & \lambda_1 v^2 ,\nonumber\\
m_{H^\pm}^2 & \ = \ & \mu_2^2 + \frac{1}{2}\lambda_3 v^2 , \nonumber\\
m_{H^0}^2 & \ = \ & \mu_2^2 + \frac{1}{2}(\lambda_3+\lambda_4+\lambda_5)v^2 \ = \ m^2_{H^\pm}+
\frac{1}{2}\left(\lambda_4+\lambda_5\right)v^2  , \nonumber\\
m_{A^0}^2 & \ = \ & \mu_2^2 + \frac{1}{2}(\lambda_3+\lambda_4-\lambda_5)v^2 \ = \ m^2_{H^\pm}+
\frac{1}{2}\left(\lambda_4-\lambda_5\right)v^2 \, .
\label{mass_relation}
\end{eqnarray}
 Without any loss of generality, we consider $\lambda_5 <0, \lambda_4+\lambda_5 <0$ so that the CP-even scalar is the lightest $Z_2$ odd particle and hence a stable DM candidate.

\section{Neutrino Mass} \label{sec:nu} 

\begin{figure}[t!]
\centering
\includegraphics[scale=0.75]{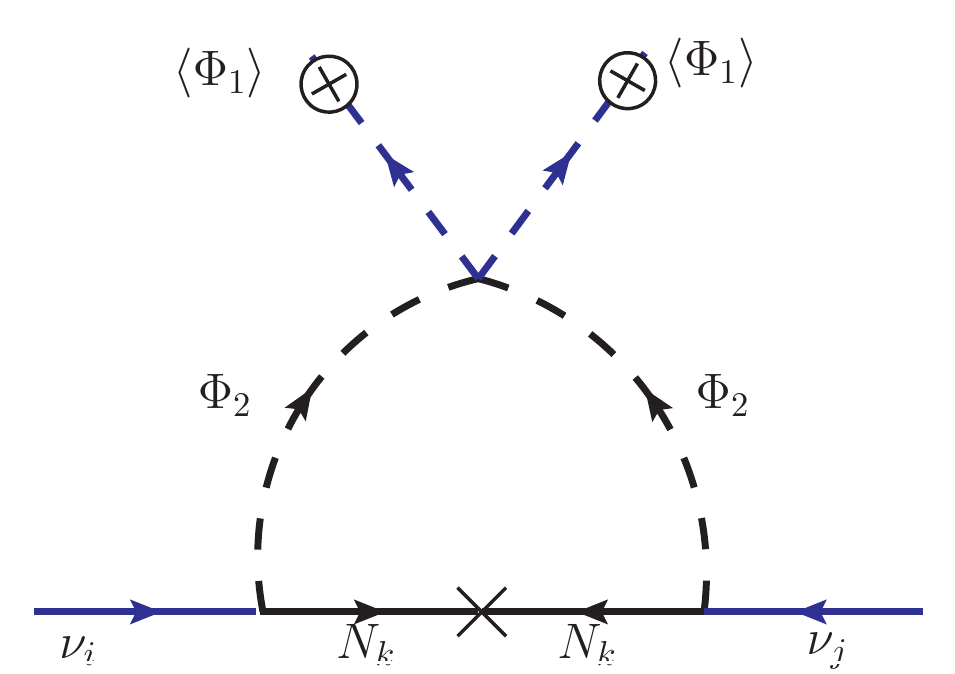}
\caption{\textit{One-loop contribution to neutrino mass in the scotogenic model.}}
\label{fig2a}
\end{figure}

The Lagrangian involving the newly added $Z_2$ odd SM-singlet fermions is
\begin{equation}\label{IRHYukawa}
{\cal L} \ \supset \ \frac{1}{2}(M_N)_{ij} N_iN_j + \left(Y_{ij} \, \bar{L}_i \tilde{\Phi}_2 N_j  + \text{H.c.} \right) \ . 
\end{equation}
Note that the $Z_2$ symmetry~\eqref{eq:Z2} forbids the usual Dirac Yukawa term $\bar{L}\tilde{\Phi}_1 N$ involving the SM Higgs, and hence, the Dirac mass term in the seesaw mechanism. So the active neutrinos remain massless at tree level. However, they can acquire a tiny mass at one-loop level through the diagram shown in Fig. \ref{fig2a}, which yields~\cite{Ma:2006km, Merle:2015ica}
\begin{align}
(M_{\nu})_{ij} \ & = \ \sum_k \frac{Y_{ik}Y_{jk} M_{k}}{32 \pi^2} \left ( \frac{m^2_{H^0}}{m^2_{H^0}-M^2_k} \: \text{ln} \frac{m^2_{H^0}}{M^2_k}-\frac{m^2_{A^0}}{m^2_{A^0}-M^2_k}\: \text{ln} \frac{m^2_{A^0}}{M^2_k} \right) \nonumber \\ 
& \ \equiv  \ \sum_k \frac{Y_{ik}Y_{jk} M_{k}}{32 \pi^2} \left[L_k(m^2_{H^0})-L_k(m^2_{A^0})\right] \, ,
\label{numass1}
\end{align}
where 
$M_k$ is the mass eigenvalue of the mass eigenstate $N_k$ in the internal line and the indices $i, j = 1,2,3$ run over the three neutrino generations as well as three copies of $N_i$. The function $L_k(m^2)$ is defined as 
\begin{align}
L_k(m^2) \ = \ \frac{m^2}{m^2-M^2_k} \: \text{ln} \frac{m^2}{M^2_k} \, .
\label{eq:Lk}
\end{align}
From Eqs.~\eqref{mass_relation}, we note that $m^2_{H^0}-m^2_{A^0}=\lambda_5 v^2$. It implies that the neutral components of the inert doublet become mass-degenerate in the limit $\lambda_5\to 0$. In this limit, the light neutrinos masses also vanish [cf.~Eq.~\eqref{numass1}], as expected, since the $\lambda_5$-term in the scalar potential~\eqref{c} breaks lepton number by two units, when considered together with the SM-singlet fermions Lagrangian~\eqref{IRHYukawa}. Therefore, the smallness of $\lambda_5$ is technically natural in the 't Hooft sense~\cite{tHooft:1979rat}.  


In this model, we need to choose some of the Yukawa couplings of the order of $10^{-3}-10^{-4}$ for ${\cal O}(10~\textrm{TeV})$ $N_i$ masses and for $\lambda_5 \sim 10^{-4}$ (typical values needed to satisfy the baryogenesis constraints; see Sec.~\ref{sec5}). One must make sure that the light neutrino masses obtained from a given choice of Yukawa couplings is consistent with the cosmological limit on the sum of neutrino masses, $\sum_i m_{i}\leq 0.11$ eV~\cite{Aghanim:2018eyx}, as well as the neutrino oscillation data~\cite{Esteban:2016qun}. This can be ensured by working in the Casas-Ibarra parametrization~\cite{Casas:2001sr}. For this purpose, we rewrite Eq.~\eqref{numass1} in a form resembling the type-I seesaw formula: 
\begin{align}
M_\nu \ = \ Y \widetilde{M}^{-1} Y^T \, ,
\label{eq:nu2}
\end{align}
where we have introduced the diagonal matrix $\widetilde{M}$ with elements
\begin{align}
 \widetilde{M}_i \ & = \ \frac{2\pi^2}{\lambda_5}\zeta_i\frac{2M_i}{v^2} \, , \\
\textrm {and}\quad \zeta_i & \ = \  \left(\frac{M_{i}^2}{8(m_{H^0}^2-m_{A^0}^2)}\left[L_i(m_{H^0}^2)-L_i(m_{A^0}^2) \right]\right)^{-1} \, . \label{eq:zeta}
\end{align}
The light neutrino mass matrix~\eqref{eq:nu2} is diagonalized by the usual PMNS mixing matrix $U$, which is determined from the neutrino oscillation data (up to the Majorana phases): 
\begin{align}
D_\nu \ = \ U^\dag M_\nu U^* \ = \ \textrm{diag}(m_1,m_2,m_3) \, .
\end{align}   
Then the Yukawa coupling matrix satisfying the neutrino data can be written as
\begin{align}
Y \ = \ U D_\nu^{1/2} O \widetilde{M}^{1/2} \, ,
\label{eq:Yuk}
\end{align}
where $O$ is an arbitrary complex orthogonal matrix.

\section{Inflation with Inert Higgs Doublet}
\label{sec2}
The IHDM needs to be coupled nonminimally to gravity for successful inflation. The relevant action during inflation in the Jordan frame can be written as~\cite{Choubey:2017hsq} (apart from the couplings to the SM-singlet fermions) 
\begin{align}\label{action}
S \ = \ \int d^4x \sqrt{-g}\left[-\frac{1}{2}M_{\rm Pl}^2 R-D_{\mu}\Phi_1 D^{\mu}\Phi_1^\dagger-D_{\mu}\Phi_2 D^\mu \Phi_2^\dagger-V(\Phi_1,\Phi_2)-\xi_1\Phi_1^2 R-\xi_2\Phi_2^2 R\right],
\end{align}
where $g$ is the metric term in the $(-,+,+,+)$ convention, $D$ stands for the covariant derivative containing couplings with the gauge bosons which just reduces to the normal derivative $D_\mu\rightarrow \partial_\mu$ (since during inflation, there are no fields other than the inflaton), $M_{\rm Pl}$ is the reduced Planck mass, $R$ is the Ricci scalar, and $\xi_1$, $\xi_2$ are dimensionless couplings of the Higgs doublets to gravity.\footnote{Quantum effects invariably give rise to such couplings at the Planck scale~\cite{Birrell:1982ix}.} To have inflation along the inert doublet direction and not the SM Higgs doublet direction, we need  $\frac{\lambda_2}{\xi_2^2}\ll \frac{\lambda_1}{\xi_1^2}$, which is satisfied for large $\xi_2$ compared to $\lambda_2$. 

To make the calculations easier, we make a conformal transformation of the metric to $\tilde{g}_{\mu\nu}=\Omega^2(x)g_{\mu\nu}$ and transform the action to the Einstein frame~\cite{Capozziello:1996xg, Kaiser:2010ps} where it looks like a regular field theory action with no explicit couplings to gravity. Without going into the details, we just quote the result for the redefined potential~\cite{Choubey:2017hsq}: 
\begin{eqnarray}\label{potential}
V_e & \ \simeq \ & \frac{\lambda_2 M_{\rm Pl}^4}{4\xi_2^2}\left[1-\exp\left(-\sqrt{\frac{2}{3}}\frac{X}{M_{\rm Pl}}\right)\right]^2 \, , \\ 
{\rm where} \quad X& \ = \ &\sqrt{\frac{3}{2}}M_{\rm Pl}\log\left(\Omega^2\right) \, .\label{defA}
\end{eqnarray}
The potential in Eq.~\eqref{potential} belongs to the Starobinsky class~\cite{Starobinsky:1979ty}; see also Ref.~\cite{Calmet:2016fsr}. The potential is almost flat at high field values $X\gg M_{\rm Pl}$, ensuring slow-roll of the inflaton field. The slow-roll parameters are given by 
\begin{align}
\epsilon & \ = \ \frac{1}{2}M_{\rm Pl}^2\left(\frac{V'_e}{V_e}\right)^2 \ = \ \frac{4}{3}
\frac{1}{\left[-1+\exp\left(\sqrt{\frac{2}{3}}\frac{X}{M_{\rm Pl}}\right)\right]^{2}} \, , \label{eq:epsilon} \\
\eta & \ = \ M_{\rm Pl}^2\left(\frac{V''_e}{V_e}\right) \ = \ 
\frac{4}{3}\frac{2-\exp\left(\sqrt{\frac{2}{3}}\frac{X}{M_{\rm Pl}}\right)}{\left[-1+\exp\left(\sqrt{\frac{2}{3}}\frac{X}{M_{\rm Pl}}\right)\right]^2} \, , \label{eq:eta}
\end{align}
where $V'_e\equiv dV_e/dX$ and $V''_e\equiv d^2V_e/dX^2$. During the inflationary epoch, $\epsilon,\eta \ll 1$ and inflation ends when $\epsilon\simeq 1$. Using this in Eq.~\eqref{eq:epsilon}, we obtain the field value $X_{\rm end}$ at the end of inflation to be 
\begin{align}
\sqrt{\frac{2}{3}}\frac{X_{\rm end}}{M_{\rm Pl}} \ \simeq \ 0.77 \, .
\label{eq:Xend}
\end{align} 
The initial field value $X_{\rm in}$ at the beginning of inflation can then be obtained from the number of $e$-folds, $N$ (the number of times the Universe expanded by $e$ times its own size), defined as 
\begin{align} 
N \ & = \ \frac{1}{M^2_{\rm Pl}}\int\limits_{X_{\rm end}}^{X_{\rm in}}\frac{V_e}{V'_e}dX \, . \label{eq:e-fold}
\end{align}
In principle, any $N>50$ could solve the flatness, horizon and entropy problems of the standard Big Bang cosmology~\cite{Remmen:2014mia}. We choose $N=60$ which is typically expected if the energy scale of inflation is of the order of $10^{16}$ GeV (GUT scale)~\cite{Liddle:2003as}. Using this value and Eq.~\eqref{eq:Xend} in Eq.~\eqref{eq:e-fold}, we obtain 
\begin{align}
\sqrt{\frac{2}{3}}\frac{X_{\rm in}}{M_{\rm Pl}} \ \simeq \ 4.45 \, .
\label{eq:Xin}
\end{align} 

The slow-roll parameters are approximately constant during slow-roll inflation, because 
\begin{align}
\frac{d\epsilon}{dN} \ \simeq \ 2\epsilon(\eta-2\epsilon) \ = \ {\cal O}(\epsilon^2) \, .
\end{align} 
So we can evaluate the inflationary parameters like the tensor-to-scalar ratio $r$ and spectral index ($n_s$) from Eqs.~\eqref{eq:epsilon} and \eqref{eq:eta} using the initial field value $X_{\rm in}$ from Eq.~\eqref{eq:Xin}. We obtain
 \begin{align}
 r \ & = \ 16\,\epsilon \ = \ 0.0029 \, , \\
n_s \ & = \ 1-6\epsilon+2\eta \ = \ 0.9678 \, , 
\end{align}
which are consistent with the 2018 Planck constraints (Planck TT,TE, EE+lowE+lensing)~\cite{Akrami:2018odb}:
\begin{align}
r  & \ < \ 0.11 & & (\text{at 95\% C.L.})\, ,\\
n_s & \ = \  0.9649\pm 0.0042 & & (\text{at 68\% C.L.}) \, .
\end{align} 
Similarly, the scalar power spectrum amplitude can be estimated as  
\begin{eqnarray}
P_s \ = \ \frac{1}{12\,\pi^2}\frac{V_e^3}{M_{\rm Pl}^6\,V_e^{'2}} \ \simeq \ 5.57\:\frac{\lambda_2}{\xi_2^2} \, .
\end{eqnarray}
Using the 2018 Planck result for $\log(10^{10}P_s)=3.047\pm 0.014$ at 68\% C.L.~\cite{Akrami:2018odb}, we find a relation between $\lambda_2$ and $\xi_2$:
\begin{eqnarray}
\xi_2 \ \simeq \ 5.33\times 10^4 \,\lambda_2^{1/2} \, .
\label{eq:xi2}
\end{eqnarray}
As we will see in the following section, successful reheating after inflation imposes a lower bound on $\lambda_2$, which in turn puts a lower bound on $\xi_2$ by virtue of Eq.~\eqref{eq:xi2}. 

We also note that in the IHDM, although we have two complex scalar fields during inflation, only the inert doublet components contribute to the effective potential given by Eq.~\eqref{potential}. Thus, the isocurvature fluctuations typically present in multi-field inflation models are suppressed here. To be specific, the isocurvature fraction is predicted to be $\beta_{\rm iso}\sim {\cal O}(10^{-5})$~\cite{Choubey:2017hsq}, which is consistent with the Planck constraints.   

\section{Reheating}
\label{sec3}
At the end of inflation, the energy density stored in the inflaton field starts to disperse as the inflaton annihilates or decays into other particles, including those of the SM. This is the reheating epoch~\cite{Allahverdi:2010xz}, which takes the Universe from the matter-dominated phase during inflation to the radiation-domination phase.




As $X$ falls below $M_{\rm Pl}$, the inflationary potential in Eq.~\eqref{potential} can be approximated by a quadratic potential: 
\begin{equation}
V_e \ \simeq \ \frac{\lambda_2M_{\rm Pl}^2}{6\xi_2^2} \, X^2 \ \equiv \ \frac{1}{2}\omega^2 X^2 \, , \qquad \text{where}\quad \omega^2=\frac{\lambda_2M_{\rm Pl}^2}{3\xi_2^2} \, .\label{PotRt}
\end{equation}
Reheating occurs in this harmonic oscillator potential well~\cite{Linde:1981mu} where the field $X$ undergoes very rapid coherent oscillations with frequency $\omega$. The equation of motion for $X$ during reheating is 
\begin{equation}
\ddot{X}+3H\dot{X}+\frac{dV_e}{dX} \ = \ 0 \, ,
\label{eq:Xev}
\end{equation}
where $\dot{X}\equiv dX/dt$, $\ddot{X}\equiv d^2X/dt^2$ and $H$ is the Hubble expansion rate. On solving Eq.~\eqref{eq:Xev} for $\omega\gg H$, we obtain
\begin{equation}
X \ = \ X_0(t)\cos (\omega t) \, , \qquad \text{with} \quad X_0(t) \ = \ 2\sqrt{2}\frac{\xi_2}{\sqrt{\lambda_2}}\frac{1}{t} \, .
\end{equation}
We define the time $t_{\rm cr}=\frac{2\xi_2}{\omega}$ at which the amplitude $X_0$ crosses $X_{\rm cr} =\sqrt{\frac{2}{3}}\frac{M_{\rm Pl}}{\xi_2}$, which marks the end of reheating.

In the IHDM inflation, where the inert doublet is the inflaton, it can decay into the $W$ and $Z$ bosons through the kinetic coupling term $\frac{g^2}{4\sqrt{6}}\frac{M_{\rm Pl}}{\xi_2} X\, W^2$ and into the SM Higgs boson through the quartic coupling terms $\lambda_i\sqrt{\frac{2}{3}}\frac{M_{\rm Pl}}{\xi_2} X|\Phi_1|^2$ (where $i=3,4,5$ in the scalar potential~\eqref{c}).  The SM particles do not have a physical mass at the time of reheating but acquire an effective mass due to the couplings to inflaton and its oscillations. For $\omega\gg H$, the amplitude $X_0$ can be taken to be constant over one oscillation period. This allows us to write down the effective mass term for vector and scalar bosons as
\begin{eqnarray}
m_W^2& \ = \ &\frac{g^2}{2\sqrt{6}}\frac{M_{\rm Pl}}{\xi_2}|X| \, ,\\
m_h^2 & \ = \ &\frac{1}{\sqrt{6}}\left(\lambda_3+\frac{\lambda_4}{2}\right)\frac{M_{\rm Pl}}{\xi_2}|X| \, .
\end{eqnarray}
The effective coupling of the $W$ boson is large enough that it is produced as a nonrelativistic species. The same is true for the Higgs boson, if any of the $\lambda_i$'s above are of order 1. So the decay and annihilation of these bosons to relativistic SM fermions will reheat the Universe.  

Following Ref.~\cite{Bezrukov:2008ut} (see also Refs.~\cite{GarciaBellido:2008ab, Repond:2016sol}), the production of $W$ and Higgs bosons in the linear and resonance regions are, respectively, given by 
\begin{align}\label{Wprod}
\frac{d(n_W a^3)}{dt} \ & = \ \left\{\begin{array}{c} \frac{P}{2\pi^3}\omega K_1^3 a^3,\;\;\;\;  \textrm{(linear)}, \\ 2a^3\omega Q n_W, \;\;\;\;  \textrm{(resonance)},\end{array}\right. \\
\frac{d(n_ha^3)}{dt} \ & = \ \left\{\begin{array}{c} \frac{P}{2\pi^3}\omega K_2^3 a^3,\;\;\;\;  \textrm{(linear)}, \\ 2 a^3 \omega Q n_h.\;\;\;\; \textrm{(resonance)} , 
\end{array}\right. \label{hprod}
\end{align}
where $P$ and $Q$ are numerical factors with $P\approx 0.0455$ and $Q\approx 0.045$, $\alpha_W=\frac{g^2}{4\pi}$ is the weak coupling constant, $a$ is the scale factor, $n_{W},n_h$ are the number densities of $W$ and Higgs, respectively, and 
\begin{eqnarray}
K_1& \ = \ &\left[\frac{g^2 M_{\rm Pl}^2}{6\xi_2^2}\sqrt{\frac{\lambda_2}{2}}\: X_0(t_i)\right]^{1/3}, \\
K_2 & \ = \ & \left[\frac{\left(\lambda_3+\frac{\lambda_4}{2}\right)\,M_{\rm Pl}^2}{3\xi_2^2}\sqrt{\frac{\lambda_2}{2}}\: X_0(t_i)\right]^{1/3},
\end{eqnarray}
where $t_i$ is the instant when the inflaton field value $X=0$. Inflaton can decay into $W$ and Higgs bosons only in the vicinity of $X=0$ when the effective masses of these bosons are much smaller than the inflaton effective mass $\omega$.

At low number densities $n_W$ and $n_h$ of the produced $W$ and Higgs bosons, their decays to SM fermions are the dominant channels for the production of relativistic particles and successful reheating of the Universe to the radiation-dominated epoch. If the number densities become large, their production rates will increase exponentially due to parametric resonance during which the bosons will mostly annihilate to produce fermions. Fermion production through decay of $W$ takes place very slowly and would reheat the Universe long after the resonance period has ended~\cite{Bezrukov:2008ut}, while production through annihilation is a much faster process. Since annihilation can only occur when the number density is large enough, this necessitates the occurrence of parametric resonance.  

Parametric resonance production of $W$ bosons can occur only when the $W$ boson decay rate, given by 
\begin{equation} \label{Wdecay}
\Gamma_W \ = \ \frac{3}{4} \: \alpha_W m_W \, ,
\end{equation}
falls below its resonance production rate given by Eq.~\eqref{Wprod}. This leads to the condition 
\begin{equation}\label{WresC}
X_0 \ \lesssim \ \frac{3.56}{\pi}\frac{Q^2\,\lambda_2}{\alpha_W^3}X_{\rm cr} \ \approx \ 61.88\,\lambda_2\,X_{\rm cr} \, ,
\end{equation}
which imposes a {\it lower} bound on $\lambda_2$: 
\begin{equation}
\lambda_2 \ \gtrsim \ \frac{1}{60} \, .\label{eq:lamb2}
\end{equation}

As for the resonance production of Higgs bosons, this occurs when the decay rate of Higgs into fermions, governed by the Yukawa couplings $y_f$, given by 
\begin{equation}\label{hdecay}
\Gamma_h \ = \ \frac{y_f^2}{16\pi}m_h \, ,
\end{equation}
falls below its resonance production rate given by Eq.~\eqref{hprod}. Keeping only the largest Yukawa coupling $y_t$ to top quarks in Eq.~\eqref{hdecay}, we get the Higgs resonance condition 
\begin{equation}\label{hresC}
X_0 \ \lesssim \ \frac{64\pi\,Q^2\lambda_2}{(\lambda_3+\frac{\lambda_4}{2})\,y_t^4}X_{\rm cr} \ \approx \ 0.41 \left(\frac{\lambda_2}{\lambda_3+\frac{\lambda_4}{2}}\right)\,X_{\rm cr} \, ,
\end{equation}
Comparing Eq.~\eqref{hresC} to Eq.~\eqref{WresC}, we find that the Higgs production will enter the parametric resonance regime around the same time as the $W$ boson, only if $\lambda_3+\frac{\lambda_4}{2}\lesssim 0.006$. Moreover, from Eq.~\eqref{hresC}, we find that for Higgs resonance production to occur, $\lambda_3+\frac{\lambda_4}{2}\lesssim 0.41\lambda_2$. On the other hand, since the neutral component of the inert doublet is also the DM candidate in our case, we need its couplings to the SM Higgs of order 1 (see Sec.~\ref{sec4} below). Therefore, the Higgs production will not enter resonance regime till long after the end of the quadratic phase of the potential. Thus, the production rate of the Higgs remains small and can be neglected, as compared to the $W$ boson production, as far as the reheating is concerned in our IHDM scenario.

Once the $W$ boson production has entered the parametric resonance regime, they rapidly annihilate to transfer their entire energy density to relativistic fermions (radiation), given by~\cite{Choubey:2017hsq} 
\begin{equation}
\rho_r \ \simeq \ \frac{1.06 \times 10^{57}~\text{GeV}^4}{\lambda_2} \, .
\label{eq:rhor1}
\end{equation}

The inert doublet also couples to the $Z_2$ odd SM-singlet fermions and SM leptons [cf.~Eq.~\eqref{IRHYukawa}], and, therefore, can directly decay into them during reheating (as long as the effective inflaton mass $\omega$ is larger than the $N_k$ mass $M_k$) though the effective coupling terms 
\begin{equation}
-Y_{ij} \sqrt{\frac{M_{\rm Pl}X}{\sqrt{24}}} (\bar{e}_i N_j + \bar{N}_j e_i)+Y_{ij}\sqrt{\frac{M_{\rm Pl}X}{\sqrt{24}}}(\bar{\nu}_i N_j+\bar{N}_j \nu_i) \, ,
\end{equation}
where $i,j$ are the lepton flavor indices. Taking a representative value of $Y_{ij}=Y\approx 10^{-4}$, we get for the energy density
\begin{equation}
\rho_r \ = \ \sqrt{\frac{3}{\lambda_2}} \frac{Y^2 M_{\rm Pl} \omega^3}{4\pi} \ \simeq \  \frac{6.16\times 10^{49}~\text{GeV}^4}{\sqrt{\lambda_2}} \, ,
\label{eq:rhor2}
\end{equation}
which is much smaller compared to the relativistic energy density produced by gauge bosons [cf.~Eq.~\eqref{eq:rhor1}], if their parametric resonance production occurs.


Using Eq.~\eqref{eq:rhor1} and taking $\lambda_2 \sim {\cal O}(1)$ (see Sec.~\ref{sec6}), we can compute the reheating temperature: 
\begin{equation}
T_r \ \simeq \ \left(\frac{30\,\rho_r}{\pi^2\,g_*}\right)^{1/4} \ \approx \ 10^{14}\,\textrm{GeV}\;
\label{eq:reheat}
\end{equation}
where $g_* = 116$ is the number of degrees of freedom in the relativistic plasma that includes the SM particles plus three $Z_2$ odd SM-singlet fermions and four additional Higgs bosons in the scotogenic model.

\section{Dark Matter}
\label{sec4}
After reheating, the remaining inert doublet particles become a part of the thermal plasma and go into thermal equilibrium until freeze-out later where the lightest of the two neutral scalars $H^0$ and $A^0$ in Eq.~\eqref{eq:idm} becomes a viable DM candidate. To obtain the allowed parameter space from the observed DM relic density considerations, we solve the Boltzmann equation for the evolution of the DM number density $n_{\rm DM}$:
\begin{equation}
\frac{dn_{\rm DM}}{dt}+3Hn_{\rm DM} \ = \ -\langle \sigma v \rangle \left[n^2_{\rm DM} -(n^{\rm eq}_{\rm DM})^2\right],
\label{eq:BE}
\end{equation}
where $n^{\rm eq}_{\rm DM}$ is the equilibrium number density of DM and $ \langle \sigma v \rangle $ is the thermally averaged annihilation cross section, given by~\cite{Gondolo:1990dk}
\begin{equation}
\langle \sigma v \rangle \ = \ \frac{1}{8m_{\rm DM}^4T K^2_2\left(\frac{m_{\rm DM}}{T}\right)} \int\limits^{\infty}_{4m_{\rm DM}^2}\sigma (s-4m_{\rm DM}^2)\sqrt{s}\: K_1\left(\frac{\sqrt{s}}{T}\right) ds \, ,
\label{eq:sigmav}
\end{equation}
where $K_i(x)$'s are modified Bessel functions of order $i$. One can solve Eq.~\eqref{eq:BE} to obtain the freeze-out temperature $T_f$ and the relic abundance $\Omega_{\rm DM}=\frac{\rho_{\rm DM}}{\rho_c}$, where $\rho_{\rm DM}$ is the DM energy density and $\rho_c=\frac{3H_0^2}{8\pi G_N}$ is the critical energy density of the Universe, with $G_N$ being Newton's gravitational constant and $H_0\equiv 100\: h~\text{km\: s}^{-1}\:\text{Mpc}^{-1}$ is the present-day Hubble expansion rate. Assuming only $s$-wave annihilation, the relic density is given by~\cite{Kolb:1990vq}
\begin{align}
\Omega_{\rm DM}h^2 \ = \ \left(1.07\times 10^9~{\rm GeV}^{-1}\right)\frac{ x_fg_*^{1/2}}{g_{*s}M_{\rm Pl}\langle \sigma v\rangle_f} \, ,
\label{eq:omega}
\end{align}
where $g_*$ and $g_{*s}$ are the effective relativistic degrees of freedom that contribute to the energy density and entropy density, respectively,\footnote{For most of the history of the Universe, all particle species had a common temperature and $g_{*s}$ can be replaced by $g_*$, for which we will use the SM value of 106.75, since the extra Higgs and $Z_2$ odd SM-singlet fermions fields are heavier than the DM and do not contribute to $g_*$ for the relic density calculation (but do contribute for the reheating calculation in the previous section).} and 
\begin{equation}
x_f \equiv \frac{m_{\rm DM}}{T_f} \ = \ \ln \left(0.038\frac{g}{g_*^{1/2}}M_{\text{Pl}}m_{\rm DM}\langle \sigma v\rangle_f\right) \, , 
\label{xf}
\end{equation} 
with $g$ being the number of internal degrees of freedom of the DM and the subscript $f$ on $\langle \sigma v \rangle$ meaning that Eq.~\eqref{eq:sigmav} is evaluated at the freeze-out temperature, which by itself is derived from the equality condition of DM interaction rate $\Gamma = n_{\rm DM} \langle \sigma v \rangle$ with the rate of expansion of the Universe $H(T) \simeq \sqrt{\frac{\pi^2 g_*}{90}}\frac{T^2}{M_{\rm Pl}}$ (i.e, the freeze-out condition).







We consider the CP-even neutral component $H^0$ of the inert scalar doublet $\Phi_2$ in Eq.~\eqref{eq:idm} as the DM candidate for our analysis. This is similar to the inert doublet model of DM discussed extensively in the literature~\cite{Ma:2006km, Cirelli:2005uq,  Barbieri:2006dq, Ma:2006fn, LopezHonorez:2006gr,  Hambye:2009pw, Dolle:2009fn, Honorez:2010re, LopezHonorez:2010tb, Gustafsson:2012aj, Goudelis:2013uca, Arhrib:2013ela, Diaz:2015pyv, Dasgupta:2014hha}. At the tree level, the annihilation of $H^0$ proceeds via the quartic interactions ($\lambda_{3,4,5}$ terms) in the scalar potential~\eqref{c}, as well as via gauge interactions with the SM $W$ and $Z$ bosons. In the low mass regime ($m_{H^0} \equiv m_{\text{DM}}\leq m_W$), the annihilation of DM to the SM fermions through s-channel Higgs mediation dominates over other channels. As pointed out in Ref.~\cite{Honorez:2010re}, the annihilation: $H^0 H^0 \rightarrow W W^* \rightarrow W f \bar{f^{\prime}}$ also plays a role in the $m_{\text{DM}} \leq m_W$ region. 
Also, when the heavier components of the inert scalar doublet have masses close to the DM mass, they can be thermally accessible at the epoch of DM freeze-out. Therefore, the annihilation cross section of DM in such a case gets additional contributions from coannihilations between the DM and the heavier components of the scalar doublet $\Phi_2$. 

In the presence of coannihilation, the effective cross section at freeze-out can be expressed as~\cite{Griest:1990kh}
\begin{align}
\sigma_{\rm eff} 
& \ = \ \sum_{i,j}^{N}\langle \sigma_{ij}v\rangle \frac{g_ig_j}{g^2_{\rm eff}}(1+\Delta_i)^{3/2}(1+\Delta_j)^{3/2}e^{-x_F(\Delta_i + \Delta_j)} \, , 
\end{align}
where $\Delta_i = \frac{m_i-m_{\text{DM}}}{m_{\text{DM}}}$ is the relative mass difference between the heavier component $i$ of the inert Higgs doublet (with $g_i$ internal degrees of freedom) and the DM,  
\begin{align}
g_{\rm eff} & \ = \ \sum_{i=1}^{N}g_i(1+\Delta_i)^{3/2}e^{-x_f\Delta_i} 
\end{align}
is the total effective degrees of freedom, and 
\begin{align}
\langle \sigma_{ij} v \rangle & \ = \ \frac{x_f}{8m^2_im^2_jm_{\text{DM}}K_2\left(\frac{m_ix_f}{m_{\text{DM}}}\right)K_2\left(\frac{m_j x_f}{m_{\text{DM}}}\right)} \nonumber \\
& \qquad \times \int\limits^{\infty}_{(m_i+m_j)^2}ds \: \sigma_{ij}\left(s-2(m_i^2+m_j^2)\right) \sqrt{s}\: K_1\left(\frac{\sqrt{s}\: x_f}{m_{\text{DM}}}\right) 
\label{eq:thcs}
\end{align}
is the modified thermally averaged cross section, compared to Eq.~\eqref{eq:sigmav}. the relic density formula~\eqref{eq:omega} gets modified to 
\begin{equation}
\Omega_{\rm DM} h^2 \ = \ \frac{2.4\times 10^{-10}}{\sigma_{\rm eff}}~\rm GeV^{-2} \, .
\end{equation}

In the present model, we include the coannihilation effects from the CP-odd scalar $A^0$ and the charged scalars $H^\pm$. The corresponding mass splittings $\Delta_{A^0}$ and  $\Delta_{H^\pm}$ depend on the values of the quartic couplings $\lambda_4$ and $\lambda_5$ [cf.~Eqs.~\eqref{mass_relation}].
In the presence of SM-singlet fermions, there exists additional annihilation and coannihilation channels which, in principle, could affect the DM relic abundance. However, in our case, they are considered to be heavy for successful vanilla leptogenesis which we discuss in the next section. For DM mass in the TeV range, such heavy fermions do not affect their relic abundance~\cite{Borah:2017dqx}.

\begin{figure}[t!]
\centering
\includegraphics[scale=1]{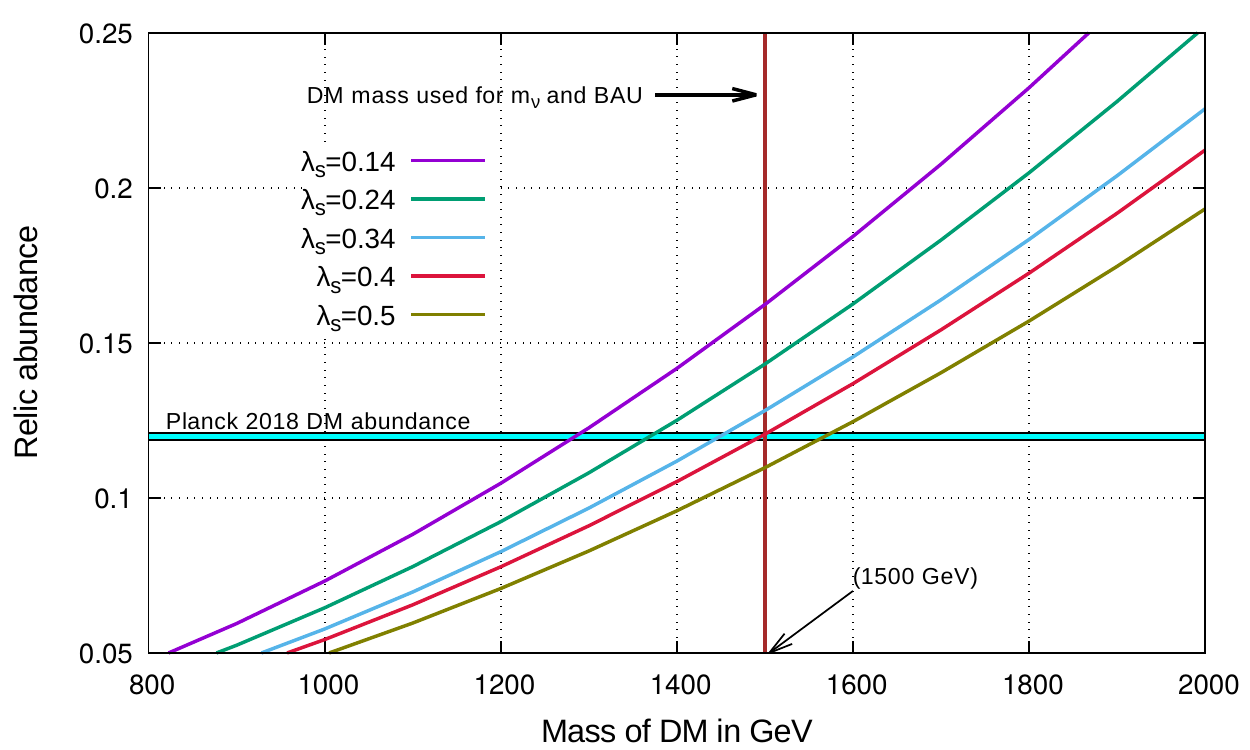}
\caption{\textit{The DM relic abundance in the scotogenic model. Increasing $\lambda_s\equiv \lambda_3+\lambda_4+\lambda_5$ beyond $\simeq 0.5$ violates the perturbative bounds at scales where inflation and reheating occur. The horizontal band is the 68\% C.L. observed DM relic abundance from Planck 2018 data~\cite{Aghanim:2018eyx}. The vertical line shows a benchmark value of the DM mass chosen for our subsequent analysis.}}
\label{fig:dm}
\end{figure}

Fig. \ref{fig:dm} shows the relic abundance of DM as a function of its mass. The different curves are for different values of $\lambda_s=\lambda_3+\lambda_4+\lambda_5$. We find that the relic abundance is satisfied for TeV scale DM, if we want the freeze-out of DM to occur in the temperature range of the EW phase transition.\footnote{This follows from the general expectation that for a WIMP DM, $x_f$ is typically in the range of 20--30.} Since we are using the same inert doublet particles to explain inflation which occurs near the Planck scale, we need to make sure that the coupling parameters $\lambda_i$ do not run beyond their perturbative limit of $\sqrt{4\pi}$ (see Sec.~\ref{sec6}). Because of this, $\lambda_3$ gets an upper bound of around 0.5 during freeze-out. Similarly, $\lambda_5$ gets constrained by baryogenesis (see Sec.~\ref{sec5}) and together these give an upper bound on $\lambda_s$ of around 0.5 (assuming $\lambda_4\simeq \lambda_5$). The observed relic abundance $\Omega_{\rm DM}^{\rm obs} h^2=0.120\pm 0.001$ at 68\% C.L.~\cite{Aghanim:2018eyx}, shown by the horizontal shaded region, together with the perturbative limit on the quartic couplings gives an {\it upper} bound on the DM mass of around 1.6 TeV. The results that we have obtained here give a spin-independent scattering cross section of $\simeq 10^{-45}-10^{-46}$ cm$^2$ and are within reach of the next generation DM direct detection experiments, like LZ~\cite{Akerib:2015cja}, XENONnT~\cite{Aprile:2015uzo}, DARWIN~\cite{Aalbers:2016jon} and PandaX-30T~\cite{Liu:2017drf}.

\section{Baryogenesis}
\label{sec5}
At the end of the inflationary era, any preexisting asymmetry between baryons and antibaryons would be exponentially suppressed and we must freshly generate the observed baryon asymmetry dynamically. This can done in the scotogenic model via the mechanism of leptogenesis by virtue of the out-of-equilibrium decay of the $N_i$~\cite{Ma:2006fn, Kashiwase:2012xd, Kashiwase:2013uy, Racker:2013lua, Clarke:2015hta, Hugle:2018qbw}. In the vanilla leptogenesis scenario with hierarchical $N_i$ masses~\cite{Buchmuller:2004nz}, there exists an absolute lower bound on the mass of the lightest $Z_2$ odd SM-singlet fermions of $M_{1}^{\rm min}\gtrsim 10^9$ GeV~\cite{Davidson:2002qv, Buchmuller:2002rq}.\footnote{Including flavor and thermal effects could, in principle, lower this bound to about $10^6$ GeV~\cite{Moffat:2018wke}.} A similar lower bound can be derived in the scotogenic model with only two $Z_2$ odd SM-singlet fermions in the strong washout regime. However, with three such SM-singlet fermions, the bound can be lowered to about 10 TeV~\cite{Hugle:2018qbw}, even without resorting to a resonant enhancement of the CP-asymmetry~\cite{Pilaftsis:2003gt, Dev:2017wwc}. We will use a hierarchical setup for the $N_i$ masses with benchmark values of $M_{1}=10$ TeV, $M_{2}=50$ TeV and $M_{3}=100$ TeV to derive the leptogenesis constraints on the model parameter space and their compatibility with the inflation and DM constraints discussed in the previous sections.

After being thermally produced during reheating, the $N_i$ in our setup start decaying into SM leptons and the inert Higgs doublet, if kinematically allowed, through their Yukawa interactions [cf.~Eq.~\eqref{IRHYukawa}]. For a hierarchical scenario with $M_{1}\ll M_{2},\, M_{3}$, the lepton asymmetry produced by decays of $N_{2,3}$ are negligible for the final lepton asymmetry, as they become suppressed due to the strong washout effects caused by $N_1$ or $N_{2,3}$-mediated interactions. Only the asymmetry created by $N_1$ decays is relevant for the generation of the final lepton asymmetry which is later converted into the baryon asymmetry of the Universe by the electro-weak sphaleron phase transitions. To obtain results for baryogenesis, we need to solve the simultaneous Boltzmann equations for $N_1$ decay and $N_{B-L}$ (the $B-L$ number density) formation. Any such $B-L$ calculation depends on a comparison between the Hubble parameter and the decay rates for $N_1\rightarrow\ell\Phi_2$, $\bar\ell\Phi_2^*$ processes which will be responsible for the asymmetry, as well as to the CP-asymmetry parameter $\varepsilon_1$. To this effect, 
we define the decay parameter
\begin{equation}
K_{N_1} \ = \ \frac{\Gamma_{N_1}}{H(z=1)} \, ,
\label{eq:Kdecay}
\end{equation}
where $\Gamma_{N_1}$ is the total decay rate of $N_1$ and $H(z=1)$ is the Hubble rate evaluated at  $z=\frac{M_{1}}{T}=1$. With our choice of the $N_i$ and DM masses, we stay in the weak washout regime ($K_{N_1}<1$), which is crucial for allowing successful leptogenesis with $M_{1}=10$ TeV~\cite{Hugle:2018qbw}. With the Yukawa couplings given in Eq.~\eqref{eq:Yuk}, the $N_1$ decay rate is given by 
\begin{eqnarray}
\Gamma_{N_1} & \ = \ & \frac{M_{1}}{8\pi}(Y^\dagger Y)_{11}\left[1-\left(\frac{m_{\rm DM}}{M_{1}}\right)^2\right]^2 \ \equiv \ \frac{M_{1}}{8\pi}(Y^\dagger Y)_{11}\left(1-\eta_1\right)^2 \, .
\label{eq:gamma1}
\end{eqnarray}
Ignoring flavor effects and summing over all flavors, the CP asymmetry parameter is given by
\begin{eqnarray}
&& \varepsilon_1  \ = \  \frac{1}{8\pi(Y^\dagger Y)_{11}}\sum_{j\neq 1}\textrm{Im}\left[(Y^\dagger Y)_{1j}^2\right] \left[f(r_{j1},\eta_1)-\frac{\sqrt{r_{j1}}}{r_{j1}-1}(1-\eta_1)^2\right] \, , \label{eq:vareps}\\
\textrm{where}
\quad && f(r_{j1},\eta_1) \ = \  \sqrt{r_{j1}}\left[1+\frac{1-2\eta_1+r_{j1}}{(1-\eta_1)^2}\;\ln\left(\frac{r_{j1}-\eta_1^2}{1-2\eta_1+r_{j1}}\right)\right] \, ,
\end{eqnarray}
and $r_{j1}  = \left(\frac{M_{j}}{M_{1}}\right)^2$, $\eta_1\equiv \left(\frac{m_{\rm DM}}{M_{1}}\right)^2$. 

\begin{figure}[t!]
\subfigure{\includegraphics[width=0.49\textwidth,keepaspectratio]{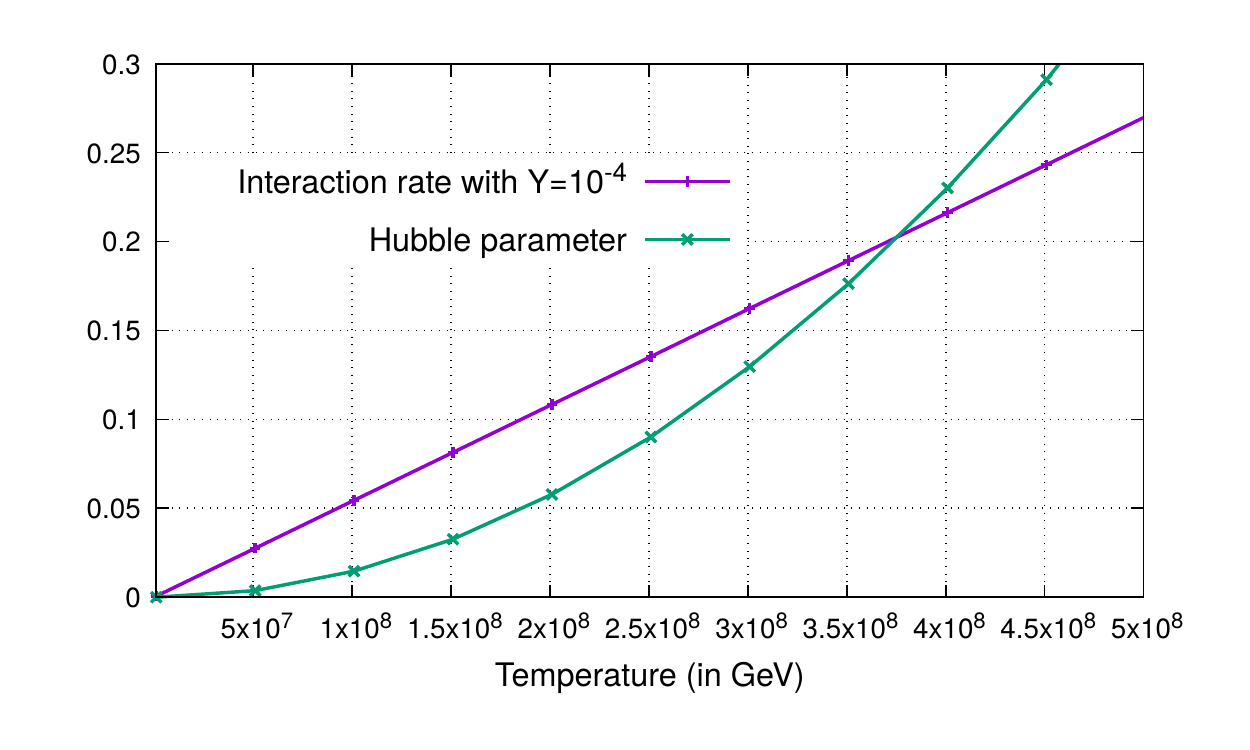}}
\subfigure{\includegraphics[width=0.49\textwidth,keepaspectratio]{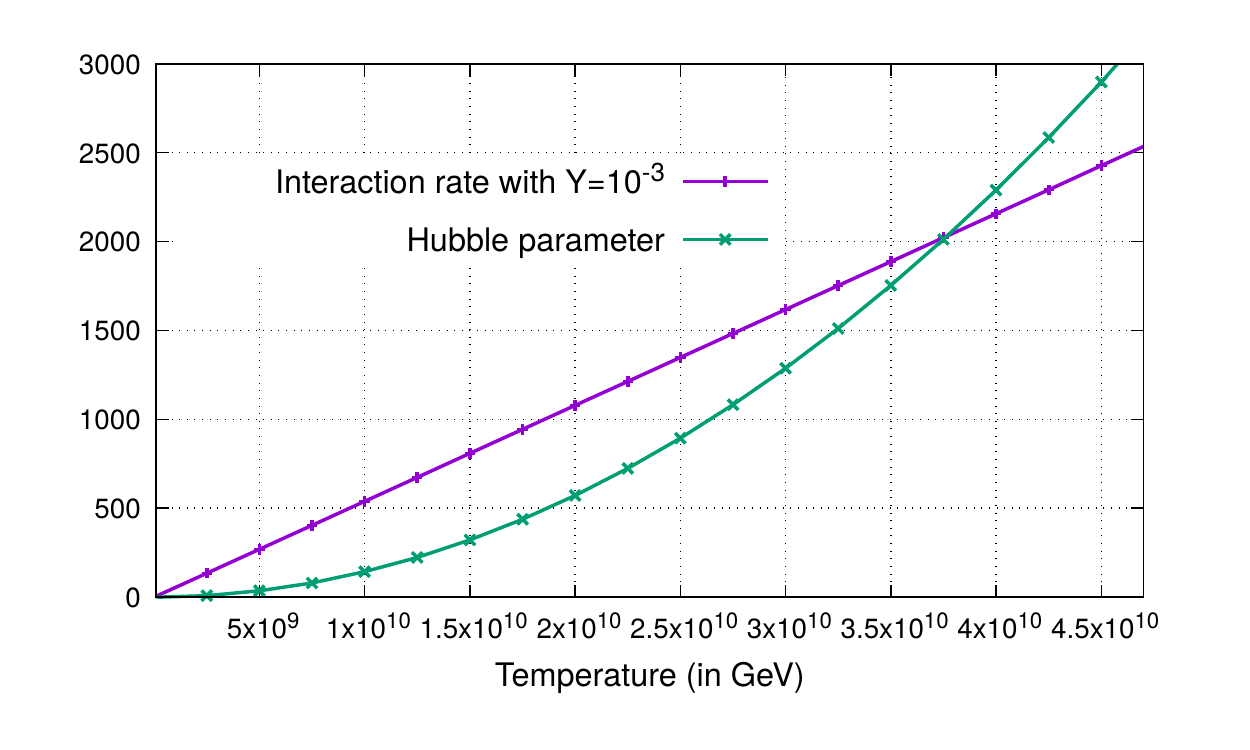}}
\caption{\textit{The interaction rate is less than the Hubble rate in the early Universe but it overtakes the Hubble rate below ${\cal O}(10^{8}~\textrm{GeV})$ for $Y=10^{-4}$ (left) and ${\cal O}(10^{10}~\textrm{GeV})$ for $Y=10^{-3}$ (right). Here we have taken $M_{1}=10$ TeV.}}
\label{fig:RateComp}
\end{figure}

To solve the Boltzmann equations, we need to start with a thermal initial abundance for $N_1$ where the interaction rate of the $N_1$ particles is above the Hubble rate after reheating. If the Yukawa couplings corresponding to $N_1$ are very small, this is not possible. This constrains us to have at least some of the $N_1$ couplings of the order of $10^{-3}-10^{-4}$. With these values, the processes $W/Z,H^{\pm}/H_0/A_0\rightarrow N,\ell^\pm/\nu$ help to satisfy an initial thermal abundance of $N_1$. Fig. \ref{fig:RateComp} shows the comparison between the interaction rate and the Hubble rate for a Yukawa coupling strength of $10^{-4}$ (left) and $10^{-3}$ (right). We can see that for a Yukawa strength of $10^{-4}$, $N_1$ particles enter the thermal plasma at temperatures of ${\cal O}(10^8~\textrm{GeV})$ while for $10^{-3}$ coupling strength, they enter the equilibrium much earlier at temperatures of ${\cal O}(10^{10}~\textrm{GeV})$. This allows us to start with an initial thermal abundance of $N_1$ while solving the Boltzmann equations given that mass of $N_1$ is taken 10 TeV below which the processes $W/Z,H^{\pm}/H_0/A_0\rightarrow N,\ell^\pm/\nu$ die out.

We numerically solve the coupled Boltzmann equations for the $N_1$ and ${B-L}$ number densities, respectively given by~\cite{Buchmuller:2004nz}
\begin{eqnarray}
\frac{dn_{N_1}}{dz}& \ = \ &-D_1 (n_{N_1}-n_{N_1}^{\rm eq}) \, , \label{eq:bol1} \\
\frac{dn_{B-L}}{dz}& \ = \ &-\varepsilon_1 D_1 (n_{N_1}-n_{N_1}^{\rm eq})-W_1 n_{B-L} \, , \label{eq:bol2}
\end{eqnarray}
where $n_{N_1}^{\rm eq}=\frac{z^2}{2}K_2(z)$ is the equilibrium number density of $N_1$ (with $K_i(z)$ being the modified Bessel function of $i$-th kind), 
\begin{align}
D_1 \ \equiv \ \frac{\Gamma_{1}}{Hz} \ = \ K_{N_1} z \frac{K_1(z)}{K_2(z)}
\end{align}
 measures the total decay rate with respect to the Hubble rate, and similarly, $W_1\equiv \frac{\Gamma_{W}}{Hz}$ measures the total washout rate. The washout term is the sum of two contributions, i.e. $W_1=W_{\rm ID}+W_{\Delta L=2}$, where the washout due to the inverse decays $\ell\Phi_2$, $\bar\ell\Phi_2^* \rightarrow N_1$ is given by 
\begin{align}
W_{\rm ID} \ = \ \frac{1}{4}K_{N_1} z^3 K_1(z) \, ,
\end{align}
 and that due to the $\Delta L=2$ scatterings $\ell\Phi_2\leftrightarrow \bar{\ell}\Phi_2^*$, $\ell\ell\leftrightarrow \Phi_2^*\Phi_2^*$ is given by~\cite{Hugle:2018qbw} 
\begin{align}
W_{\Delta L=2} \ \simeq \ \frac{18\sqrt{10}\,M_{\rm Pl}}{\pi^4 g_\ell \sqrt{g_*}z^2 v^4}\left(\frac{2\pi^2}{\lambda_5}\right)^2 M_{1}\bar{m}_\zeta^2 \, ,
\label{eq:wash2}
\end{align}
where we have assumed $\eta_1\ll 1$ for simplicity, $g_\ell$ stands for the internal degrees of freedom for the SM leptons, and  $\bar{m}_\zeta$ is the effective neutrino mass parameter, defined as 
\begin{align}
\bar{m}_\zeta^2 \  \simeq \  4\zeta_1^2 m_{1}^2+\zeta_2 m_{2}^2+\zeta_3^2 m_{3}^2 \, ,
\end{align}
with $m_i$'s being the light neutrino mass eigenvalues, $\zeta_i$ defined in Eq.~\eqref{eq:zeta} and $L_i(m^2)$ defined in Eq.~\eqref{eq:Lk}. Note that Eq.~\eqref{eq:wash2} is similar to the $\Delta L=2$ wash-out term in vanilla leptogenesis, except for the $\left(\frac{2\pi^2}{\lambda_5}\right)^2$ factor.

\begin{figure}[t!]
\centering
\includegraphics[scale=1]{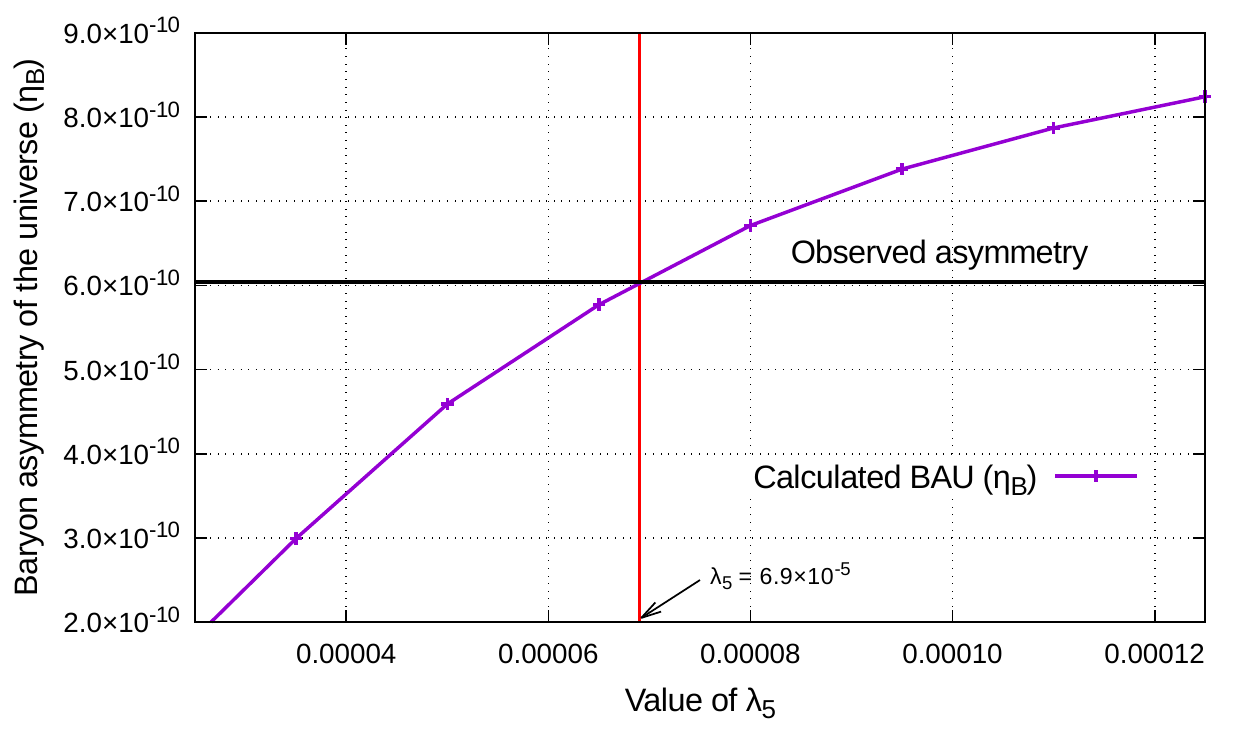}
\caption{\textit{The baryon-to-photon ratio as a function of $|\lambda_5|$ for benchmark DM mass of 1.5 TeV. The horizontal line gives the observed value, as inferred from Planck 2018 data~\cite{Aghanim:2018eyx}.}}
\label{fig:BAU}
\end{figure}

After solving the Boltzmann equations~\eqref{eq:bol1} and \eqref{eq:bol2} numerically, we evaluate the final $B-L$ asymmetry $n_{B-L}^f$ just before sphaleron freeze-out, which is then converted to the baryon-to-photon ratio 
\begin{align}
\eta_B \ = \ \frac{3}{4}\frac{g_*^{0}}{g_*}a_{\rm sph}n_{B-L}^f \ \simeq \ 9.2\times 10^{-3}\: n_{B-L}^f \, ,
\label{eq:etaB}
\end{align}
where $a_{\rm sph}=\frac{8}{23}$ is the sphaleron conversion factor (taking into account two Higgs doublets), $g_*=110.75$ is the effective relativistic degrees of freedom during the production of the final lepton asymmetry (taking into account two Higgs doublets, but not the $N_i$, since they have already frozen-out by this time), and $g_*^0=\frac{43}{11}$ is the effective relativistic degrees of freedom at the recombination epoch.  Note that the baryon asymmetry in this scenario only depends on $\lambda_5$ and is independent of other quartic couplings in the potential Eq.~\eqref{c}.  For illustration, we show in Fig.~\ref{fig:BAU} the resulting value of $\eta_B$ obtained from Eq.~\eqref{eq:etaB} as a function of the quartic coupling $\lambda_5$ for a benchmark value of $m_{H^0}=1.5$ TeV (which is also the DM mass in this setup). The horizontal line shows the observed value $\eta_B^{\rm obs}=(6.04\pm 0.08)\times 10^{-10}$ at 68\% C.L., as inferred from the Planck 2018 data on the baryon density $\Omega_Bh^2=0.0224\pm 0.0001$~\cite{Aghanim:2018eyx}. For our choice of the $N_i$ masses $M_{1}=10$ TeV, $M_{2}=50$ TeV and $M_{3}=100$ TeV, the Yukawa matrix satisfying the light neutrino data can be obtained from Eq.~\eqref{eq:Yuk} for a suitable choice of the orthogonal matrix $O$. In particular, we find that for successful leptogenesis, the Yukawa couplings of the lightest $Z_2$ odd SM-singlet fermion, $N_1$, are required to be of order ${\cal O}(10^{-8}-10^{-9})$, while those of $N_2$ and $N_3$ are of order ${\cal O}(10^{-4}-10^{-3})$. This is because we are in the weak washout regime, which requires $K_{N_1}<1$, and hence, smaller $(Y^\dag Y)_{11}$ [cf.~Eq.~\eqref{eq:gamma1}]. However, this also means a smaller source term for the $B-L$ asymmetry in Eq.~\eqref{eq:bol2}, which must be compensated by a larger $\varepsilon_1$, thereby requiring larger $(Y^\dag Y)_{1j}$ (with $j=2,3$) in Eq.~\eqref{eq:vareps}. 
 


\begin{figure}[t!]
\centering
\includegraphics[scale=1]{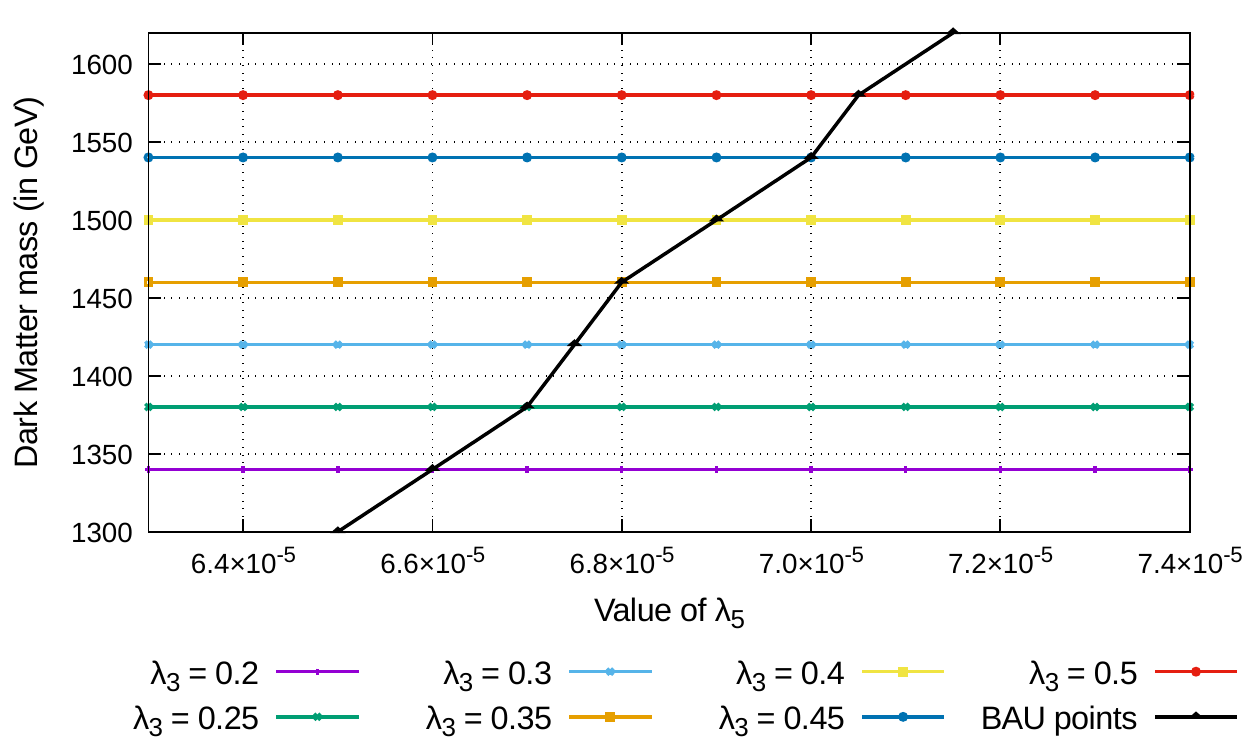}
\caption{\textit{DM mass as a function of $\lambda_3$ and $|\lambda_5|$ satisfying the relic density constraint. The black line across the graph shows the points that satisfy the baryogenesis constraint}}
\label{fig:MdmL5}
\end{figure}

In order to see the interplay of the DM and baryogenesis constraints on the model, we examine the variation of the DM mass with $|\lambda_5|$ for different values of $\lambda_3$ that satisfy the relic density constraint (assuming $\lambda_4=\lambda_5$ in $\lambda_s=\lambda_3+\lambda_4+\lambda_5$ that goes into relic calculation; see Sec.~\ref{sec4}). This allows us to get a range of DM masses and $\lambda_5$ values that allow for successful baryogenesis in the present model as is shown in Fig.~\ref{fig:MdmL5}, where the black line across the graph shows the points that satisfy the baryogenesis constraint. The lightest active neutrino mass in this model is of $\mathcal{O}(10^{-12}~\textrm{eV}$). From this result, we obtain a preferred range of 1.3--1.60 TeV for the DM mass.   



\section{Renormalization Group analysis}
\label{sec6}
Since we are combining inflation, which is a Planck-scale phenomenon, with baryogenesis and DM physics at the TeV-scale, we must make sure that all the coupling values used in this analysis remain perturbative up to the Planck scale. The requirements among the couplings for inflation and reheating is that the inflaton self-coupling $\lambda_2$ be greater than around $1/60$ to successfully reheat the Universe at temperatures of the order of $10^{13}-10^{14}$ GeV. On the other hand, the inflaton is also the dark matter candidate with TeV range mass. This will freeze-out around the electroweak scales only if the quartic couplings of the inert doublet to the Higgs are of $\mathcal{O}(1)$. However, such large couplings at EW scales run the risk of exceeding the perturbative bound of $4\pi$. This constrains the available Higgs-inert doublet couplings depending on the values of $\lambda_1$ and $\lambda_2$. Therefore it becomes necessary to run the couplings from the EW scales to upto the Planck scales to keep them perturbative at low scales and high scales.

Here we show the results of the one-loop RG evolution for a typical set of coupling values used in our numerical calculations above.  
Following Refs.~\cite{Branco:2011iw, Chakrabarty:2015yia}, the one-loop RG equations for the gauge couplings of the IHDM are given by
\begin{eqnarray}
16\pi^2 \frac{dg_s}{dt} & \ = \ & - 7 g_s^3,\\
16\pi^2 \frac{dg}{dt} & \ = \ & - 3 g^3,\\
16\pi^2 \frac{dg^{\prime}}{dt} & \ = \ & 7 {g^\prime}^3.
\end{eqnarray}
where $g^{\prime}$, $g$ and $g_s$ denote the U$(1)$, SU$(2)_L$ and SU$(3)_c$ gauge couplings respectively, and $t\equiv \ln(\mu)$ is the energy scale. 
The quartic couplings $ \lambda_{i} ~(i=1,\ldots,5)$ evolve as follows: 
\begin{eqnarray}
16\pi^2 \frac{d\lambda_1}{dt} &\ = \ &
12 \lambda_1^2 + 4 \lambda_3^2 + 4 \lambda_3 \lambda_4 + 2 \lambda_4^2
+ 2  \lambda_5^2 + \frac{3}{4}(3g^4 + g^{\prime 4} +2 g^2 g^{\prime 2})\nonumber \\
 & & -\lambda_1 (9 g^2 + 3 g^{\prime
2} - 12 y_t^2 - 12 y_b^2 - 4 y_{\tau}^2 ) - 12 y_t^4\,, \\
16\pi^2 \frac{d\lambda_2}{dt} &\ = \ &
12 \lambda_2^2 + 4 \lambda_3^2 + 4 \lambda_3 \lambda_4 + 2 \lambda_4^2
+ 2 \lambda_5^2  \nonumber \\
 & &
+\
\frac{3}{4}(3g^4 + g^{\prime 4} +2g^2 g^{\prime 2}) -3\lambda_2
(3g^2 +g^{\prime 2}- \frac{4}{3} Y^2)- 4 Y^4\,, \\
16\pi^2 \frac{d\lambda_3}{dt}  & \ = \ &
\left( \lambda_1 + \lambda_2 \right) \left( 6 \lambda_3 + 2 \lambda_4 \right)
+ 4 \lambda_3^2 + 2 \lambda_4^2
+ 2 \lambda_5^2
+\frac{3}{4}(3g^4 + g^{\prime 4} -2g^2 g^{\prime 2}) \nonumber \\
& & - \lambda_3
(9g^2 + 3g^{\prime 2}- 6 y_t^2 - 6 y_b^2 - 2 y_{\tau}^2- 2 Y^2)\,, \\
16\pi^2 \frac{d\lambda_4}{dt}  & \ = \ &
2 \left( \lambda_1 + \lambda_2 \right) \lambda_4
+ 8 \lambda_3 \lambda_4 + 4 \lambda_4^2
+ 8 \lambda_5^2 
+\ 3g^2 g^{\prime 2} \nonumber \\
& &- \lambda_4 (9g^2 + 3g^{\prime
2}- 6 y_t^2 - 6 y_b^2 - 2 y_{\tau}^2 - 2 Y^2)\,,\\
16\pi^2 \frac{d\lambda_5}{dt}  & \ = \ &
\left( 2 \lambda_1 + 2 \lambda_2 
+ 
8 \lambda_3 + 12 \lambda_4 \right) \lambda_5
- \ \lambda_5 (9g^2 + 3g^{\prime 2} - 6 y_b^2 - 2 y_{\tau}^2- 6 y_t^2- 2 Y^2)\,,   
\end{eqnarray}

For the Yukawa couplings the corresponding set of RG equations are
\begin{eqnarray}
 16\pi^2 \frac{dy_b}{dt} & \ = \ & y_{b}\left(-8g_s^2 - \frac94 g^2 - \frac{5}{12} g^{\prime 2}+
 \frac92 y_{b}^2 +y_{\tau}^2 + \frac32 y_{t}^2\right)\,,\\
 16\pi^2 \frac{dy_t}{dt} & \ = \ & y_{t}\left(-8g_s^2 - \frac94 g^2 - \frac{17}{12} g^{\prime 2}+
 \frac92 y_{t}^2 +y_{\tau}^2 + \frac32 y_{b}^2\right)\,,\\
 16\pi^2 \frac{dy_{\tau}}{dt} & \ = \ & y_{\tau}\left(-\frac94 g^2 - \frac{15}{4} g^{\prime 2}+ 3y_{b}^2 + 3y_{t}^2 + \frac12 Y^2 + \frac52 y_{\tau}^2\right )\,.\\
  16\pi^2 \frac{dY}{dt} & \ = \ & y_{\tau}\left(-\frac94 g^2 - \frac{3}{4} g^{\prime 2} - \frac34 y_{\tau}^2 + \frac52 Y^2\right)\,.
\end{eqnarray}

\begin{figure}[t!]
\centering
\subfigure{\includegraphics[width=0.45\textwidth,keepaspectratio]{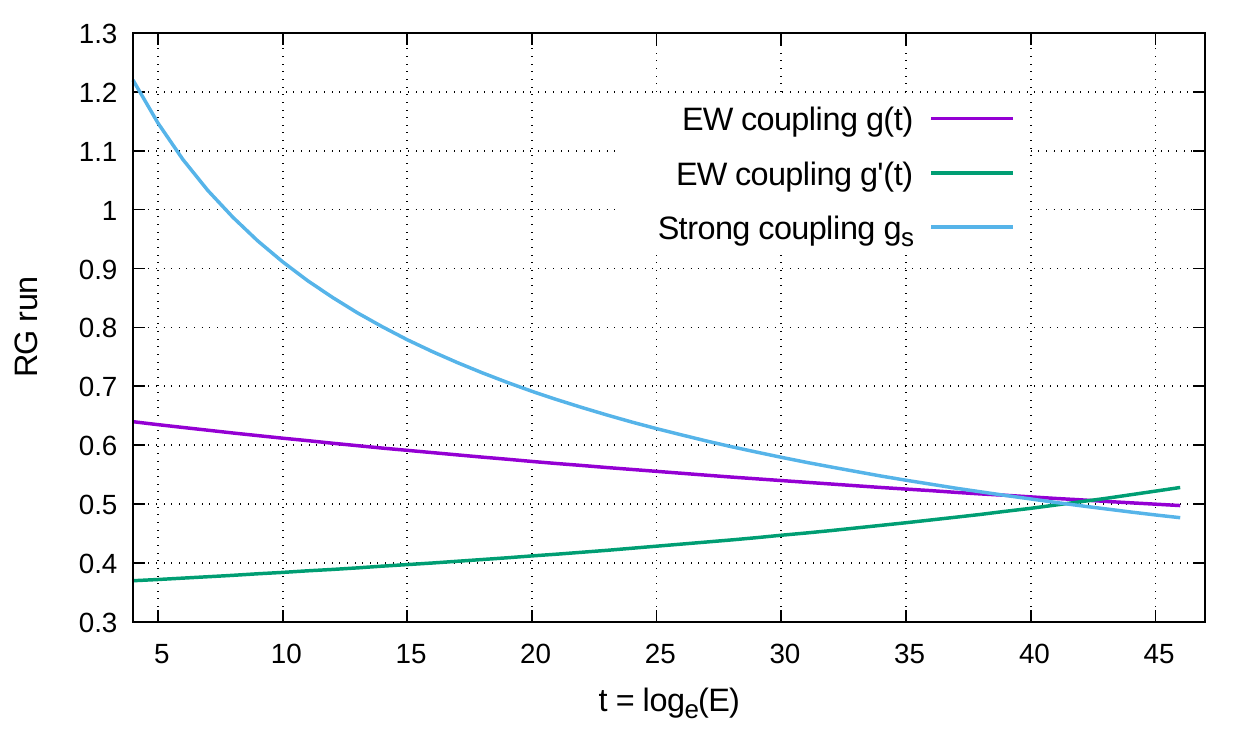}}
\subfigure{\includegraphics[width=0.45\textwidth,keepaspectratio]{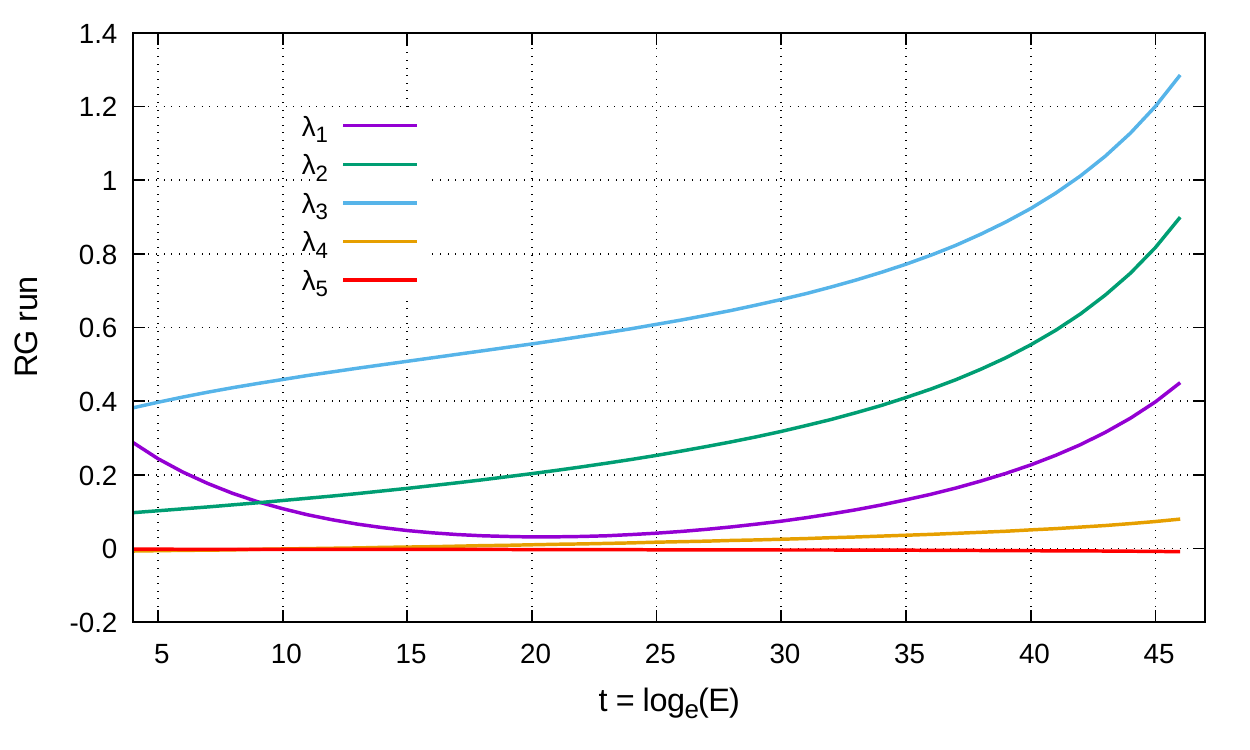}}
\subfigure{\includegraphics[width=0.45\textwidth,keepaspectratio]{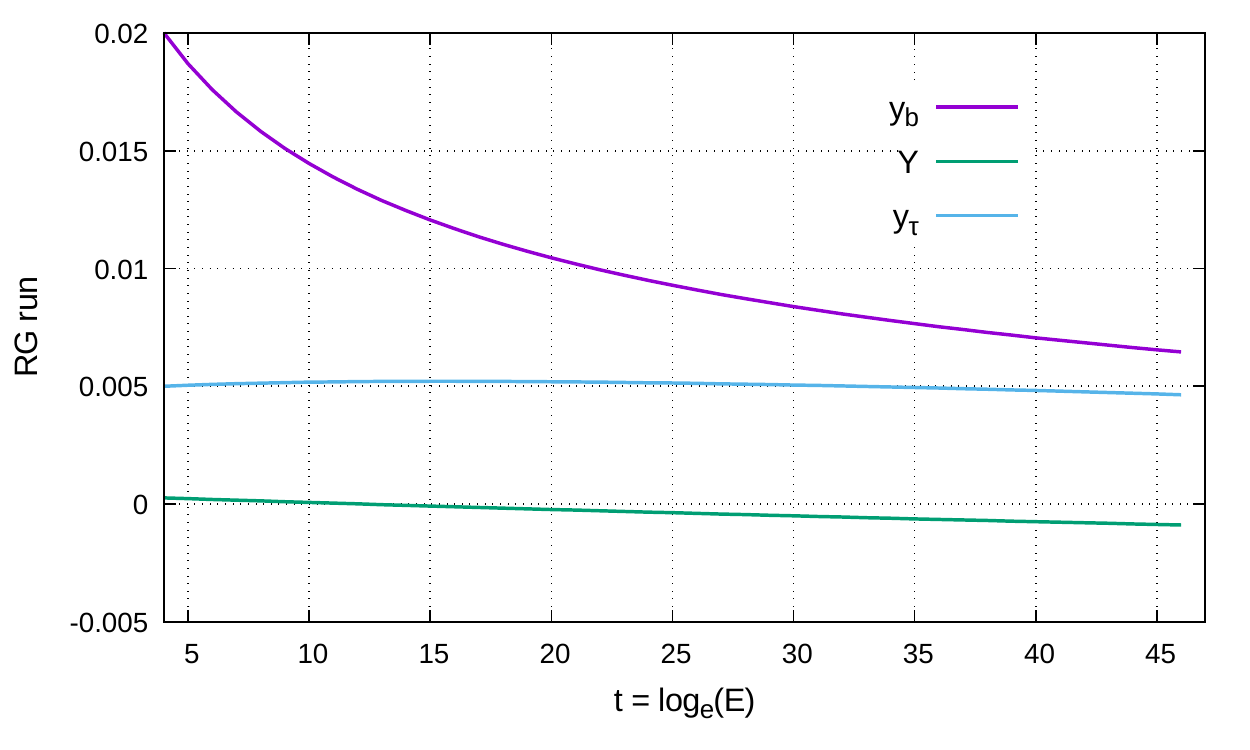}}
\subfigure{\includegraphics[width=0.45\textwidth,keepaspectratio]{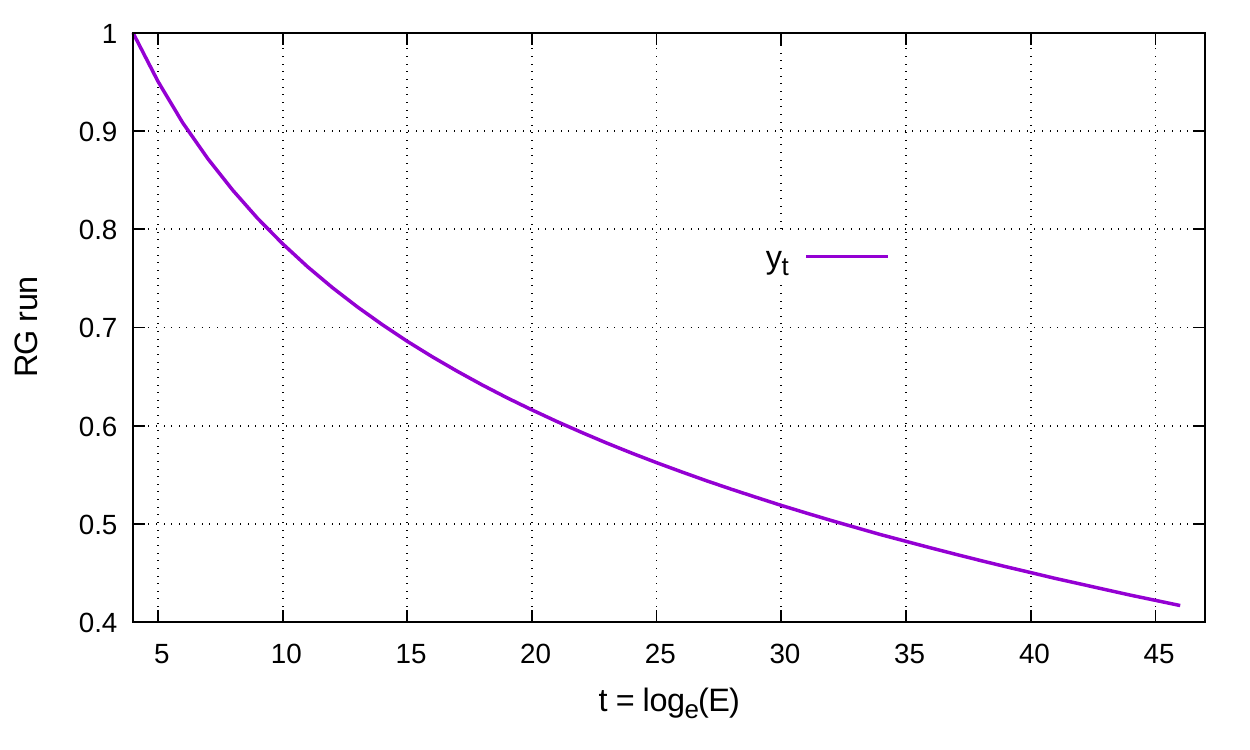}}
\caption{\textit{The RG evolution of the electroweak and strong gauge couplings (top left), scalar quartic couplings (top right), Yukawa couplings for bottom, tau and $Z_2$ odd SM-singlet fermions (bottom left) and top Yukawa coupling (bottom right) with the initial values given in the text.}}
\label{fig:RGRun}
\end{figure}

The following values of the couplings were taken as the initial values for illustration:
\begin{equation}
\begin{aligned}
\lambda_1 = 0.26;\;\;\lambda_2=0.1;\;\;\lambda_3=0.45\;\;\textrm{at the EW scale}\nonumber\\
y_t = 1;\;\;y_\tau = 0.005;\;\;y_b = 0.02\;\;\textrm{at the EW scale}\nonumber\\
g = 0.64;\;\;g^{\prime} = 0.37;\;\;g_s = 1.22\;\;\textrm{at the EW scale}\nonumber\\
\lambda_4=-0.0025;\;\;\lambda_5=-0.0025\;\;\textrm{at 10 TeV ($N_1$ mass scale)} \nonumber\\
\;\;Y = 0.0001\;\;\textrm{at 10 TeV ($N_1$ mass scale)} \nonumber
\end{aligned}
\end{equation}
The RG evolution up to the Planck scale for all these couplings is shown in  Fig.~\ref{fig:RGRun}. We conclude that for the parameter values chosen in our numerical analysis, the couplings remain perturbative all the way up to the Planck scale, thus simultaneously allowing inflation at high scale, and baryogenesis and DM freeze-out at TeV scale. 

We note that the one-loop RG equations given above are sufficient to describe the running of the couplings at the desired accuracy level. To verify this, we have numerically checked the two-loop running using {\tt SARAH}~\cite{Staub:2008uz} and find that the scalar couplings $\lambda_{1,2,3,4,5}$ run very close to their one-loop running values up to the Planck scale. Only the lepton Yukawa coupling starts differing at around $10^8$ GeV. However, around this energy, it starts decreasing for both one-loop and two-loop runs, so there is no danger of it ever crossing the perturbative bound. Thus, the use of the one-loop RG evolution equations is justified in terms of the required accuracy.

\section{Conclusion}
\label{sec7}
In this work, we have taken the inert Higgs doublet model extended by three $Z_2$ odd SM-singlet fermions as the overarching framework to successfully achieve inflation, reheating, dark matter relic density, baryogenesis, and neutrino masses. Naturally small neutrino masses are obtained via one-loop graphs involving the $N_i$ and inert doublet scalars coupling to the SM Higgs doublet. nonminimal coupling of the inert doublet (identified as the inflaton) to gravity allows us to do Starobinsky-type inflation and provides an excellent fit to the inflationary observables. The Universe is reheated after inflation by parametric resonance production of gauge bosons and $Z_2$ odd SM-singlet fermions from the annihilation of the inflaton particles. This gives us a lower bound on the quartic coupling $\lambda_2\gtrsim\frac{1}{60}$. The inert doublet particles become part of the thermal plasma and later freeze-out to provide the required DM relic density. We find that the preferred DM mass range is between 1.25--1.60 TeV. Successful baryogenesis is achieved by the decays of the $N_i$ to SM leptons and the inert doublet scalars. We obtain a small range for $|\lambda_5|$ between $6.5\times10^{-5}$ to $7.2\times 10^{-5}$ which can reproduce the observed baryon asymmetry of the Universe for the lightest $N_i$ mass scale of 10 TeV, while simultaneously satisfying the neutrino mass constraints.

There have been previous attempts at explaining inflation, dark matter, baryogenesis and neutrino masses together in a single framework, including $\nu$MSM, minimal supersymmetric standard model and axion models~\cite{Shaposhnikov:2006xi, Allahverdi:2007wt, Kohri:2009ka, Boucenna:2014uma, Salvio:2015cja, Ballesteros:2016euj}. The obvious simplicity of our model compared to supersymmetric models is the absence of supersymmetry up to the GUT scale. A model that involves supersymmetry \cite{Allahverdi:2007wt} uses a combination of Higgs, sleptons and sneutrino to act as the inflaton. The lightest sneutrino is the DM candidate and there is an extra $U(1)$ gauge boson $Z'$ and an extra gaugino $\tilde{Z}'$, which help bring the right-handed sneutrino into thermal equilibrium to obtain the correct relic density. In our model, a single scalar doublet forms the inflaton and the lightest neutral scalar component becomes the DM candidate. Another model with supersymmetry \cite{Kohri:2009ka} studies an inflection point inflation scenario in which the reheating temperature is close to the big bang nucleosynthesis temperature ($T_r\sim 1-10$ MeV). Considering such ultralow reheating temperatures removes the possibility of using electroweak sphaleron processes to create a baryon asymmetry and they need to invoke $R$-parity violating operators to directly produce the asymmetry in the baryonic sector. Gravitinos produced from inflaton decay can be long-lived enough to form the DM, as the $R$-parity violation ensures that the lightest supersymmetry particle is not stable. In contrast, in our model the $Z_2$ symmetry is unbroken and provides stability to dark matter. The model in Ref.~\cite{Boucenna:2014uma} uses the simplest type-I seesaw with three SM singlet fermions for neutrino mass generation at tree level and a complex scalar singlet carrying lepton number $L=2$. The real part of the scalar singlet gives inflation, while the complex part which is the associated Nambu-Goldstone boson (Majoron) and gets a small mass from soft lepton number breaking forms the DM. Another such minimalistic model explaining inflation, dark matter, baryogenesis and neutrino masses is the $\nu$MSM with three right-handed neutrinos for type-I seesaw, extended by a scalar singlet~\cite{Shaposhnikov:2006xi}. However, these two models use a chaotic or  natural inflation scenario, mostly motivated to explain the BICEP2 observation of a large tensor-to-scalar ratio~\cite{Ade:2014xna} -- a claim later retracted after incorporating the polarized emission from galactic dust~\cite{Ade:2015tva}. In general, these inflationary models are getting increasingly constrained by Planck observations~\cite{Akrami:2018odb}. In contrast, coupling the scalars nonminimally to gravity in our model takes it into the Starobinsky class of models which sit in the sweet-spot of inflationary observations. However, our model does not aim to explain the strong CP problem which gives an advantage to models that include axions~\cite{Salvio:2015cja, Ballesteros:2016euj}.

\acknowledgments
We thank the organizers of WHEPP XV at IISER Bhopal (December 14-23, 2017), where this work was initiated. B.D. thanks Jackson Clarke, Moritz Platscher and Kai Schmitz for useful discussions and correspondence on a related idea. D.B. acknowledges the support from IIT Guwahati start-up grant (reference number:  xPHYSUGI-ITG01152xxDB001) and Associateship Programme of IUCAA, Pune. The work of B.D. is supported by the US Department of Energy under Grant No. DE-SC0017987. A.K. would like to thank the Department of Atomic Energy (DAE) Neutrino Project under the XII plan of Harish-Chandra Research Institute. The work of A.K. was supported in part by the INFOSYS scholarship for senior students. 

\bibliographystyle{apsrev}
\bibliography{ref}

\begin{thebibliography}{112}
\expandafter\ifx\csname natexlab\endcsname\relax\def\natexlab#1{#1}\fi
\expandafter\ifx\csname bibnamefont\endcsname\relax
  \def\bibnamefont#1{#1}\fi
\expandafter\ifx\csname bibfnamefont\endcsname\relax
  \def\bibfnamefont#1{#1}\fi
\expandafter\ifx\csname citenamefont\endcsname\relax
  \def\citenamefont#1{#1}\fi
\expandafter\ifx\csname url\endcsname\relax
  \def\url#1{\texttt{#1}}\fi
\expandafter\ifx\csname urlprefix\endcsname\relax\def\urlprefix{URL }\fi
\providecommand{\bibinfo}[2]{#2}
\providecommand{\eprint}[2][]{\url{#2}}

\bibitem[{\citenamefont{Tanabashi et~al.}(2018)}]{Tanabashi:2018oca}
\bibinfo{author}{\bibfnamefont{M.}~\bibnamefont{Tanabashi}}
  \bibnamefont{et~al.} (\bibinfo{collaboration}{Particle Data Group}),
  \bibinfo{journal}{Phys. Rev.} \textbf{\bibinfo{volume}{D98}},
  \bibinfo{pages}{030001} (\bibinfo{year}{2018}).

\bibitem[{\citenamefont{Zwicky}(1933)}]{Zwicky:1933gu}
\bibinfo{author}{\bibfnamefont{F.}~\bibnamefont{Zwicky}},
  \bibinfo{journal}{Helv. Phys. Acta} \textbf{\bibinfo{volume}{6}},
  \bibinfo{pages}{110} (\bibinfo{year}{1933}), \bibinfo{note}{[Gen. Rel.
  Grav.41,207(2009)]}.

\bibitem[{\citenamefont{Rubin and Ford}(1970)}]{Rubin:1970zza}
\bibinfo{author}{\bibfnamefont{V.~C.} \bibnamefont{Rubin}} \bibnamefont{and}
  \bibinfo{author}{\bibfnamefont{W.~K.} \bibnamefont{Ford},
  \bibfnamefont{Jr.}}, \bibinfo{journal}{Astrophys. J.}
  \textbf{\bibinfo{volume}{159}}, \bibinfo{pages}{379} (\bibinfo{year}{1970}).

\bibitem[{\citenamefont{Clowe et~al.}(2006)\citenamefont{Clowe, Bradac,
  Gonzalez, Markevitch, Randall, Jones, and Zaritsky}}]{Clowe:2006eq}
\bibinfo{author}{\bibfnamefont{D.}~\bibnamefont{Clowe}},
  \bibinfo{author}{\bibfnamefont{M.}~\bibnamefont{Bradac}},
  \bibinfo{author}{\bibfnamefont{A.~H.} \bibnamefont{Gonzalez}},
  \bibinfo{author}{\bibfnamefont{M.}~\bibnamefont{Markevitch}},
  \bibinfo{author}{\bibfnamefont{S.~W.} \bibnamefont{Randall}},
  \bibinfo{author}{\bibfnamefont{C.}~\bibnamefont{Jones}}, \bibnamefont{and}
  \bibinfo{author}{\bibfnamefont{D.}~\bibnamefont{Zaritsky}},
  \bibinfo{journal}{Astrophys. J.} \textbf{\bibinfo{volume}{648}},
  \bibinfo{pages}{L109} (\bibinfo{year}{2006}), \eprint{astro-ph/0608407}.

\bibitem[{\citenamefont{Aghanim et~al.}(2018)}]{Aghanim:2018eyx}
\bibinfo{author}{\bibfnamefont{N.}~\bibnamefont{Aghanim}} \bibnamefont{et~al.}
  (\bibinfo{collaboration}{Planck}) (\bibinfo{year}{2018}),
  \eprint{1807.06209}.

\bibitem[{\citenamefont{Taoso et~al.}(2008)\citenamefont{Taoso, Bertone, and
  Masiero}}]{Taoso:2007qk}
\bibinfo{author}{\bibfnamefont{M.}~\bibnamefont{Taoso}},
  \bibinfo{author}{\bibfnamefont{G.}~\bibnamefont{Bertone}}, \bibnamefont{and}
  \bibinfo{author}{\bibfnamefont{A.}~\bibnamefont{Masiero}},
  \bibinfo{journal}{JCAP} \textbf{\bibinfo{volume}{0803}}, \bibinfo{pages}{022}
  (\bibinfo{year}{2008}), \eprint{0711.4996}.

\bibitem[{\citenamefont{Feng}(2010)}]{Feng:2010gw}
\bibinfo{author}{\bibfnamefont{J.~L.} \bibnamefont{Feng}},
  \bibinfo{journal}{Ann. Rev. Astron. Astrophys.}
  \textbf{\bibinfo{volume}{48}}, \bibinfo{pages}{495} (\bibinfo{year}{2010}),
  \eprint{1003.0904}.

\bibitem[{\citenamefont{Kolb and Turner}(1990)}]{Kolb:1990vq}
\bibinfo{author}{\bibfnamefont{E.~W.} \bibnamefont{Kolb}} \bibnamefont{and}
  \bibinfo{author}{\bibfnamefont{M.~S.} \bibnamefont{Turner}},
  \bibinfo{journal}{Front. Phys.} \textbf{\bibinfo{volume}{69}},
  \bibinfo{pages}{1} (\bibinfo{year}{1990}).

\bibitem[{\citenamefont{Sakharov}(1967)}]{Sakharov:1967dj}
\bibinfo{author}{\bibfnamefont{A.~D.} \bibnamefont{Sakharov}},
  \bibinfo{journal}{Pisma Zh. Eksp. Teor. Fiz.} \textbf{\bibinfo{volume}{5}},
  \bibinfo{pages}{32} (\bibinfo{year}{1967}), \bibinfo{note}{[Usp. Fiz.
  Nauk161,no.5,61(1991)]}.

\bibitem[{\citenamefont{Weinberg}(1979)}]{Weinberg:1979bt}
\bibinfo{author}{\bibfnamefont{S.}~\bibnamefont{Weinberg}},
  \bibinfo{journal}{Phys. Rev. Lett.} \textbf{\bibinfo{volume}{42}},
  \bibinfo{pages}{850} (\bibinfo{year}{1979}).

\bibitem[{\citenamefont{Kolb and Wolfram}(1980)}]{Kolb:1979qa}
\bibinfo{author}{\bibfnamefont{E.~W.} \bibnamefont{Kolb}} \bibnamefont{and}
  \bibinfo{author}{\bibfnamefont{S.}~\bibnamefont{Wolfram}},
  \bibinfo{journal}{Nucl. Phys.} \textbf{\bibinfo{volume}{B172}},
  \bibinfo{pages}{224} (\bibinfo{year}{1980}), \bibinfo{note}{[Erratum: Nucl.
  Phys.B195,542(1982)]}.

\bibitem[{\citenamefont{Fukugita and Yanagida}(1986)}]{Fukugita:1986hr}
\bibinfo{author}{\bibfnamefont{M.}~\bibnamefont{Fukugita}} \bibnamefont{and}
  \bibinfo{author}{\bibfnamefont{T.}~\bibnamefont{Yanagida}},
  \bibinfo{journal}{Phys. Lett.} \textbf{\bibinfo{volume}{B174}},
  \bibinfo{pages}{45} (\bibinfo{year}{1986}).

\bibitem[{\citenamefont{Kuzmin et~al.}(1985)\citenamefont{Kuzmin, Rubakov, and
  Shaposhnikov}}]{Kuzmin:1985mm}
\bibinfo{author}{\bibfnamefont{V.~A.} \bibnamefont{Kuzmin}},
  \bibinfo{author}{\bibfnamefont{V.~A.} \bibnamefont{Rubakov}},
  \bibnamefont{and} \bibinfo{author}{\bibfnamefont{M.~E.}
  \bibnamefont{Shaposhnikov}}, \bibinfo{journal}{Phys. Lett.}
  \textbf{\bibinfo{volume}{155B}}, \bibinfo{pages}{36} (\bibinfo{year}{1985}).

\bibitem[{\citenamefont{Fong et~al.}(2012)\citenamefont{Fong, Nardi, and
  Riotto}}]{Fong:2013wr}
\bibinfo{author}{\bibfnamefont{C.~S.} \bibnamefont{Fong}},
  \bibinfo{author}{\bibfnamefont{E.}~\bibnamefont{Nardi}}, \bibnamefont{and}
  \bibinfo{author}{\bibfnamefont{A.}~\bibnamefont{Riotto}},
  \bibinfo{journal}{Adv. High Energy Phys.} \textbf{\bibinfo{volume}{2012}},
  \bibinfo{pages}{158303} (\bibinfo{year}{2012}), \eprint{1301.3062}.

\bibitem[{\citenamefont{Minkowski}(1977)}]{Minkowski:1977sc}
\bibinfo{author}{\bibfnamefont{P.}~\bibnamefont{Minkowski}},
  \bibinfo{journal}{Phys. Lett.} \textbf{\bibinfo{volume}{B67}},
  \bibinfo{pages}{421} (\bibinfo{year}{1977}).

\bibitem[{\citenamefont{Mohapatra and Senjanovic}(1980)}]{Mohapatra:1979ia}
\bibinfo{author}{\bibfnamefont{R.~N.} \bibnamefont{Mohapatra}}
  \bibnamefont{and}
  \bibinfo{author}{\bibfnamefont{G.}~\bibnamefont{Senjanovic}},
  \bibinfo{journal}{Phys. Rev. Lett.} \textbf{\bibinfo{volume}{44}},
  \bibinfo{pages}{912} (\bibinfo{year}{1980}).

\bibitem[{\citenamefont{Yanagida}(1979)}]{Yanagida:1979as}
\bibinfo{author}{\bibfnamefont{T.}~\bibnamefont{Yanagida}},
  \bibinfo{journal}{Conf. Proc.} \textbf{\bibinfo{volume}{C7902131}},
  \bibinfo{pages}{95} (\bibinfo{year}{1979}).

\bibitem[{\citenamefont{Gell-Mann et~al.}(1979)\citenamefont{Gell-Mann, Ramond,
  and Slansky}}]{GellMann:1980vs}
\bibinfo{author}{\bibfnamefont{M.}~\bibnamefont{Gell-Mann}},
  \bibinfo{author}{\bibfnamefont{P.}~\bibnamefont{Ramond}}, \bibnamefont{and}
  \bibinfo{author}{\bibfnamefont{R.}~\bibnamefont{Slansky}},
  \bibinfo{journal}{Conf. Proc.} \textbf{\bibinfo{volume}{C790927}},
  \bibinfo{pages}{315} (\bibinfo{year}{1979}), \eprint{1306.4669}.

\bibitem[{\citenamefont{Glashow}(1980)}]{Glashow:1979nm}
\bibinfo{author}{\bibfnamefont{S.~L.} \bibnamefont{Glashow}},
  \bibinfo{journal}{NATO Sci. Ser. B} \textbf{\bibinfo{volume}{61}},
  \bibinfo{pages}{687} (\bibinfo{year}{1980}).

\bibitem[{\citenamefont{Schechter and Valle}(1980)}]{Schechter:1980gr}
\bibinfo{author}{\bibfnamefont{J.}~\bibnamefont{Schechter}} \bibnamefont{and}
  \bibinfo{author}{\bibfnamefont{J.~W.~F.} \bibnamefont{Valle}},
  \bibinfo{journal}{Phys. Rev.} \textbf{\bibinfo{volume}{D22}},
  \bibinfo{pages}{2227} (\bibinfo{year}{1980}).

\bibitem[{\citenamefont{Guth}(1981)}]{Guth:1980zm}
\bibinfo{author}{\bibfnamefont{A.~H.} \bibnamefont{Guth}},
  \bibinfo{journal}{Phys. Rev.} \textbf{\bibinfo{volume}{D23}},
  \bibinfo{pages}{347} (\bibinfo{year}{1981}).

\bibitem[{\citenamefont{Linde}(1982)}]{Linde:1981mu}
\bibinfo{author}{\bibfnamefont{A.~D.} \bibnamefont{Linde}},
  \bibinfo{journal}{Phys. Lett.} \textbf{\bibinfo{volume}{108B}},
  \bibinfo{pages}{389} (\bibinfo{year}{1982}).

\bibitem[{\citenamefont{Komatsu et~al.}(2011)}]{Komatsu:2010fb}
\bibinfo{author}{\bibfnamefont{E.}~\bibnamefont{Komatsu}} \bibnamefont{et~al.}
  (\bibinfo{collaboration}{WMAP}), \bibinfo{journal}{Astrophys. J. Suppl.}
  \textbf{\bibinfo{volume}{192}}, \bibinfo{pages}{18} (\bibinfo{year}{2011}),
  \eprint{1001.4538}.

\bibitem[{\citenamefont{Akrami et~al.}(2018)}]{Akrami:2018odb}
\bibinfo{author}{\bibfnamefont{Y.}~\bibnamefont{Akrami}} \bibnamefont{et~al.}
  (\bibinfo{collaboration}{Planck}) (\bibinfo{year}{2018}),
  \eprint{1807.06211}.

\bibitem[{\citenamefont{Mazumdar and Rocher}(2011)}]{Mazumdar:2010sa}
\bibinfo{author}{\bibfnamefont{A.}~\bibnamefont{Mazumdar}} \bibnamefont{and}
  \bibinfo{author}{\bibfnamefont{J.}~\bibnamefont{Rocher}},
  \bibinfo{journal}{Phys. Rept.} \textbf{\bibinfo{volume}{497}},
  \bibinfo{pages}{85} (\bibinfo{year}{2011}), \eprint{1001.0993}.

\bibitem[{\citenamefont{Linde}(1983)}]{Linde:1983gd}
\bibinfo{author}{\bibfnamefont{A.~D.} \bibnamefont{Linde}},
  \bibinfo{journal}{Phys. Lett.} \textbf{\bibinfo{volume}{129B}},
  \bibinfo{pages}{177} (\bibinfo{year}{1983}).

\bibitem[{\citenamefont{Bezrukov and Shaposhnikov}(2008)}]{Bezrukov:2007ep}
\bibinfo{author}{\bibfnamefont{F.~L.} \bibnamefont{Bezrukov}} \bibnamefont{and}
  \bibinfo{author}{\bibfnamefont{M.}~\bibnamefont{Shaposhnikov}},
  \bibinfo{journal}{Phys. Lett.} \textbf{\bibinfo{volume}{B659}},
  \bibinfo{pages}{703} (\bibinfo{year}{2008}), \eprint{0710.3755}.

\bibitem[{\citenamefont{Bezrukov et~al.}(2011)\citenamefont{Bezrukov, Magnin,
  Shaposhnikov, and Sibiryakov}}]{Bezrukov:2010jz}
\bibinfo{author}{\bibfnamefont{F.}~\bibnamefont{Bezrukov}},
  \bibinfo{author}{\bibfnamefont{A.}~\bibnamefont{Magnin}},
  \bibinfo{author}{\bibfnamefont{M.}~\bibnamefont{Shaposhnikov}},
  \bibnamefont{and}
  \bibinfo{author}{\bibfnamefont{S.}~\bibnamefont{Sibiryakov}},
  \bibinfo{journal}{JHEP} \textbf{\bibinfo{volume}{01}}, \bibinfo{pages}{016}
  (\bibinfo{year}{2011}), \eprint{1008.5157}.

\bibitem[{\citenamefont{Sher}(1989)}]{Sher:1988mj}
\bibinfo{author}{\bibfnamefont{M.}~\bibnamefont{Sher}}, \bibinfo{journal}{Phys.
  Rept.} \textbf{\bibinfo{volume}{179}}, \bibinfo{pages}{273}
  (\bibinfo{year}{1989}).

\bibitem[{\citenamefont{Lerner and McDonald}(2010)}]{Lerner:2009na}
\bibinfo{author}{\bibfnamefont{R.~N.} \bibnamefont{Lerner}} \bibnamefont{and}
  \bibinfo{author}{\bibfnamefont{J.}~\bibnamefont{McDonald}},
  \bibinfo{journal}{JCAP} \textbf{\bibinfo{volume}{1004}}, \bibinfo{pages}{015}
  (\bibinfo{year}{2010}), \eprint{0912.5463}.

\bibitem[{\citenamefont{Shaposhnikov and Tkachev}(2006)}]{Shaposhnikov:2006xi}
\bibinfo{author}{\bibfnamefont{M.}~\bibnamefont{Shaposhnikov}}
  \bibnamefont{and} \bibinfo{author}{\bibfnamefont{I.}~\bibnamefont{Tkachev}},
  \bibinfo{journal}{Phys. Lett.} \textbf{\bibinfo{volume}{B639}},
  \bibinfo{pages}{414} (\bibinfo{year}{2006}), \eprint{hep-ph/0604236}.

\bibitem[{\citenamefont{Allahverdi et~al.}(2007)\citenamefont{Allahverdi,
  Dutta, and Mazumdar}}]{Allahverdi:2007wt}
\bibinfo{author}{\bibfnamefont{R.}~\bibnamefont{Allahverdi}},
  \bibinfo{author}{\bibfnamefont{B.}~\bibnamefont{Dutta}}, \bibnamefont{and}
  \bibinfo{author}{\bibfnamefont{A.}~\bibnamefont{Mazumdar}},
  \bibinfo{journal}{Phys. Rev. Lett.} \textbf{\bibinfo{volume}{99}},
  \bibinfo{pages}{261301} (\bibinfo{year}{2007}), \eprint{0708.3983}.

\bibitem[{\citenamefont{Kohri et~al.}(2009)\citenamefont{Kohri, Mazumdar, and
  Sahu}}]{Kohri:2009ka}
\bibinfo{author}{\bibfnamefont{K.}~\bibnamefont{Kohri}},
  \bibinfo{author}{\bibfnamefont{A.}~\bibnamefont{Mazumdar}}, \bibnamefont{and}
  \bibinfo{author}{\bibfnamefont{N.}~\bibnamefont{Sahu}},
  \bibinfo{journal}{Phys. Rev.} \textbf{\bibinfo{volume}{D80}},
  \bibinfo{pages}{103504} (\bibinfo{year}{2009}), \eprint{0905.1625}.

\bibitem[{\citenamefont{Boucenna et~al.}(2014)\citenamefont{Boucenna, Morisi,
  Shafi, and Valle}}]{Boucenna:2014uma}
\bibinfo{author}{\bibfnamefont{S.~M.} \bibnamefont{Boucenna}},
  \bibinfo{author}{\bibfnamefont{S.}~\bibnamefont{Morisi}},
  \bibinfo{author}{\bibfnamefont{Q.}~\bibnamefont{Shafi}}, \bibnamefont{and}
  \bibinfo{author}{\bibfnamefont{J.~W.~F.} \bibnamefont{Valle}},
  \bibinfo{journal}{Phys. Rev.} \textbf{\bibinfo{volume}{D90}},
  \bibinfo{pages}{055023} (\bibinfo{year}{2014}), \eprint{1404.3198}.

\bibitem[{\citenamefont{Salvio}(2015)}]{Salvio:2015cja}
\bibinfo{author}{\bibfnamefont{A.}~\bibnamefont{Salvio}},
  \bibinfo{journal}{Phys. Lett.} \textbf{\bibinfo{volume}{B743}},
  \bibinfo{pages}{428} (\bibinfo{year}{2015}), \eprint{1501.03781}.

\bibitem[{\citenamefont{Ballesteros et~al.}(2017)\citenamefont{Ballesteros,
  Redondo, Ringwald, and Tamarit}}]{Ballesteros:2016euj}
\bibinfo{author}{\bibfnamefont{G.}~\bibnamefont{Ballesteros}},
  \bibinfo{author}{\bibfnamefont{J.}~\bibnamefont{Redondo}},
  \bibinfo{author}{\bibfnamefont{A.}~\bibnamefont{Ringwald}}, \bibnamefont{and}
  \bibinfo{author}{\bibfnamefont{C.}~\bibnamefont{Tamarit}},
  \bibinfo{journal}{Phys. Rev. Lett.} \textbf{\bibinfo{volume}{118}},
  \bibinfo{pages}{071802} (\bibinfo{year}{2017}), \eprint{1608.05414}.

\bibitem[{\citenamefont{Ma}(2006{\natexlab{a}})}]{Ma:2006km}
\bibinfo{author}{\bibfnamefont{E.}~\bibnamefont{Ma}}, \bibinfo{journal}{Phys.
  Rev.} \textbf{\bibinfo{volume}{D73}}, \bibinfo{pages}{077301}
  (\bibinfo{year}{2006}{\natexlab{a}}), \eprint{hep-ph/0601225}.

\bibitem[{\citenamefont{Kofman et~al.}(1994)\citenamefont{Kofman, Linde, and
  Starobinsky}}]{Kofman:1994rk}
\bibinfo{author}{\bibfnamefont{L.}~\bibnamefont{Kofman}},
  \bibinfo{author}{\bibfnamefont{A.~D.} \bibnamefont{Linde}}, \bibnamefont{and}
  \bibinfo{author}{\bibfnamefont{A.~A.} \bibnamefont{Starobinsky}},
  \bibinfo{journal}{Phys. Rev. Lett.} \textbf{\bibinfo{volume}{73}},
  \bibinfo{pages}{3195} (\bibinfo{year}{1994}), \eprint{hep-th/9405187}.

\bibitem[{\citenamefont{Kofman et~al.}(1997)\citenamefont{Kofman, Linde, and
  Starobinsky}}]{Kofman:1997yn}
\bibinfo{author}{\bibfnamefont{L.}~\bibnamefont{Kofman}},
  \bibinfo{author}{\bibfnamefont{A.~D.} \bibnamefont{Linde}}, \bibnamefont{and}
  \bibinfo{author}{\bibfnamefont{A.~A.} \bibnamefont{Starobinsky}},
  \bibinfo{journal}{Phys. Rev.} \textbf{\bibinfo{volume}{D56}},
  \bibinfo{pages}{3258} (\bibinfo{year}{1997}), \eprint{hep-ph/9704452}.

\bibitem[{\citenamefont{Liddle and Urena-Lopez}(2006)}]{Liddle:2006qz}
\bibinfo{author}{\bibfnamefont{A.~R.} \bibnamefont{Liddle}} \bibnamefont{and}
  \bibinfo{author}{\bibfnamefont{L.~A.} \bibnamefont{Urena-Lopez}},
  \bibinfo{journal}{Phys. Rev. Lett.} \textbf{\bibinfo{volume}{97}},
  \bibinfo{pages}{161301} (\bibinfo{year}{2006}), \eprint{astro-ph/0605205}.

\bibitem[{\citenamefont{Cardenas}(2007)}]{Cardenas:2007xh}
\bibinfo{author}{\bibfnamefont{V.~H.} \bibnamefont{Cardenas}},
  \bibinfo{journal}{Phys. Rev.} \textbf{\bibinfo{volume}{D75}},
  \bibinfo{pages}{083512} (\bibinfo{year}{2007}), \eprint{astro-ph/0701624}.

\bibitem[{\citenamefont{Panotopoulos}(2007)}]{Panotopoulos:2007ri}
\bibinfo{author}{\bibfnamefont{G.}~\bibnamefont{Panotopoulos}},
  \bibinfo{journal}{Phys. Rev.} \textbf{\bibinfo{volume}{D75}},
  \bibinfo{pages}{127301} (\bibinfo{year}{2007}), \eprint{0706.2237}.

\bibitem[{\citenamefont{Liddle et~al.}(2008)\citenamefont{Liddle, Pahud, and
  Urena-Lopez}}]{Liddle:2008bm}
\bibinfo{author}{\bibfnamefont{A.~R.} \bibnamefont{Liddle}},
  \bibinfo{author}{\bibfnamefont{C.}~\bibnamefont{Pahud}}, \bibnamefont{and}
  \bibinfo{author}{\bibfnamefont{L.~A.} \bibnamefont{Urena-Lopez}},
  \bibinfo{journal}{Phys. Rev.} \textbf{\bibinfo{volume}{D77}},
  \bibinfo{pages}{121301} (\bibinfo{year}{2008}), \eprint{0804.0869}.

\bibitem[{\citenamefont{Bose and Majumdar}(2009)}]{Bose:2009kc}
\bibinfo{author}{\bibfnamefont{N.}~\bibnamefont{Bose}} \bibnamefont{and}
  \bibinfo{author}{\bibfnamefont{A.~S.} \bibnamefont{Majumdar}},
  \bibinfo{journal}{Phys. Rev.} \textbf{\bibinfo{volume}{D80}},
  \bibinfo{pages}{103508} (\bibinfo{year}{2009}), \eprint{0907.2330}.

\bibitem[{\citenamefont{Lerner and McDonald}(2009)}]{Lerner:2009xg}
\bibinfo{author}{\bibfnamefont{R.~N.} \bibnamefont{Lerner}} \bibnamefont{and}
  \bibinfo{author}{\bibfnamefont{J.}~\bibnamefont{McDonald}},
  \bibinfo{journal}{Phys. Rev.} \textbf{\bibinfo{volume}{D80}},
  \bibinfo{pages}{123507} (\bibinfo{year}{2009}), \eprint{0909.0520}.

\bibitem[{\citenamefont{Okada and Shafi}(2011)}]{Okada:2010jd}
\bibinfo{author}{\bibfnamefont{N.}~\bibnamefont{Okada}} \bibnamefont{and}
  \bibinfo{author}{\bibfnamefont{Q.}~\bibnamefont{Shafi}},
  \bibinfo{journal}{Phys. Rev.} \textbf{\bibinfo{volume}{D84}},
  \bibinfo{pages}{043533} (\bibinfo{year}{2011}), \eprint{1007.1672}.

\bibitem[{\citenamefont{De-Santiago and
  Cervantes-Cota}(2011)}]{DeSantiago:2011qb}
\bibinfo{author}{\bibfnamefont{J.}~\bibnamefont{De-Santiago}} \bibnamefont{and}
  \bibinfo{author}{\bibfnamefont{J.~L.} \bibnamefont{Cervantes-Cota}},
  \bibinfo{journal}{Phys. Rev.} \textbf{\bibinfo{volume}{D83}},
  \bibinfo{pages}{063502} (\bibinfo{year}{2011}), \eprint{1102.1777}.

\bibitem[{\citenamefont{Lerner and McDonald}(2011)}]{Lerner:2011ge}
\bibinfo{author}{\bibfnamefont{R.~N.} \bibnamefont{Lerner}} \bibnamefont{and}
  \bibinfo{author}{\bibfnamefont{J.}~\bibnamefont{McDonald}},
  \bibinfo{journal}{Phys. Rev.} \textbf{\bibinfo{volume}{D83}},
  \bibinfo{pages}{123522} (\bibinfo{year}{2011}), \eprint{1104.2468}.

\bibitem[{\citenamefont{de~la Macorra}(2012)}]{delaMacorra:2012sb}
\bibinfo{author}{\bibfnamefont{A.}~\bibnamefont{de~la Macorra}},
  \bibinfo{journal}{Astropart. Phys.} \textbf{\bibinfo{volume}{35}},
  \bibinfo{pages}{478} (\bibinfo{year}{2012}), \eprint{1201.6302}.

\bibitem[{\citenamefont{Khoze}(2013)}]{Khoze:2013uia}
\bibinfo{author}{\bibfnamefont{V.~V.} \bibnamefont{Khoze}},
  \bibinfo{journal}{JHEP} \textbf{\bibinfo{volume}{11}}, \bibinfo{pages}{215}
  (\bibinfo{year}{2013}), \eprint{1308.6338}.

\bibitem[{\citenamefont{Kahlhoefer and McDonald}(2015)}]{Kahlhoefer:2015jma}
\bibinfo{author}{\bibfnamefont{F.}~\bibnamefont{Kahlhoefer}} \bibnamefont{and}
  \bibinfo{author}{\bibfnamefont{J.}~\bibnamefont{McDonald}},
  \bibinfo{journal}{JCAP} \textbf{\bibinfo{volume}{1511}}, \bibinfo{pages}{015}
  (\bibinfo{year}{2015}), \eprint{1507.03600}.

\bibitem[{\citenamefont{Bastero-Gil et~al.}(2016)\citenamefont{Bastero-Gil,
  Cerezo, and Rosa}}]{Bastero-Gil:2015lga}
\bibinfo{author}{\bibfnamefont{M.}~\bibnamefont{Bastero-Gil}},
  \bibinfo{author}{\bibfnamefont{R.}~\bibnamefont{Cerezo}}, \bibnamefont{and}
  \bibinfo{author}{\bibfnamefont{J.~G.} \bibnamefont{Rosa}},
  \bibinfo{journal}{Phys. Rev.} \textbf{\bibinfo{volume}{D93}},
  \bibinfo{pages}{103531} (\bibinfo{year}{2016}), \eprint{1501.05539}.

\bibitem[{\citenamefont{Tenkanen}(2016)}]{Tenkanen:2016twd}
\bibinfo{author}{\bibfnamefont{T.}~\bibnamefont{Tenkanen}},
  \bibinfo{journal}{JHEP} \textbf{\bibinfo{volume}{09}}, \bibinfo{pages}{049}
  (\bibinfo{year}{2016}), \eprint{1607.01379}.

\bibitem[{\citenamefont{Choubey and Kumar}(2017)}]{Choubey:2017hsq}
\bibinfo{author}{\bibfnamefont{S.}~\bibnamefont{Choubey}} \bibnamefont{and}
  \bibinfo{author}{\bibfnamefont{A.}~\bibnamefont{Kumar}},
  \bibinfo{journal}{JHEP} \textbf{\bibinfo{volume}{11}}, \bibinfo{pages}{080}
  (\bibinfo{year}{2017}), \eprint{1707.06587}.

\bibitem[{\citenamefont{Heurtier}(2017)}]{Heurtier:2017nwl}
\bibinfo{author}{\bibfnamefont{L.}~\bibnamefont{Heurtier}},
  \bibinfo{journal}{JHEP} \textbf{\bibinfo{volume}{12}}, \bibinfo{pages}{072}
  (\bibinfo{year}{2017}), \eprint{1707.08999}.

\bibitem[{\citenamefont{Hooper et~al.}(2018)\citenamefont{Hooper, Krnjaic,
  Long, and Mcdermott}}]{Hooper:2018buz}
\bibinfo{author}{\bibfnamefont{D.}~\bibnamefont{Hooper}},
  \bibinfo{author}{\bibfnamefont{G.}~\bibnamefont{Krnjaic}},
  \bibinfo{author}{\bibfnamefont{A.~J.} \bibnamefont{Long}}, \bibnamefont{and}
  \bibinfo{author}{\bibfnamefont{S.~D.} \bibnamefont{Mcdermott}}
  (\bibinfo{year}{2018}), \eprint{1807.03308}.

\bibitem[{\citenamefont{Daido et~al.}(2018)\citenamefont{Daido, Takahashi, and
  Yin}}]{Daido:2017tbr}
\bibinfo{author}{\bibfnamefont{R.}~\bibnamefont{Daido}},
  \bibinfo{author}{\bibfnamefont{F.}~\bibnamefont{Takahashi}},
  \bibnamefont{and} \bibinfo{author}{\bibfnamefont{W.}~\bibnamefont{Yin}},
  \bibinfo{journal}{JHEP} \textbf{\bibinfo{volume}{02}}, \bibinfo{pages}{104}
  (\bibinfo{year}{2018}), \eprint{1710.11107}.

\bibitem[{\citenamefont{Daido et~al.}(2017)\citenamefont{Daido, Takahashi, and
  Yin}}]{Daido:2017wwb}
\bibinfo{author}{\bibfnamefont{R.}~\bibnamefont{Daido}},
  \bibinfo{author}{\bibfnamefont{F.}~\bibnamefont{Takahashi}},
  \bibnamefont{and} \bibinfo{author}{\bibfnamefont{W.}~\bibnamefont{Yin}},
  \bibinfo{journal}{JCAP} \textbf{\bibinfo{volume}{1705}}, \bibinfo{pages}{044}
  (\bibinfo{year}{2017}), \eprint{1702.03284}.

\bibitem[{\citenamefont{Hugle et~al.}(2018)\citenamefont{Hugle, Platscher, and
  Schmitz}}]{Hugle:2018qbw}
\bibinfo{author}{\bibfnamefont{T.}~\bibnamefont{Hugle}},
  \bibinfo{author}{\bibfnamefont{M.}~\bibnamefont{Platscher}},
  \bibnamefont{and} \bibinfo{author}{\bibfnamefont{K.}~\bibnamefont{Schmitz}},
  \bibinfo{journal}{Phys. Rev.} \textbf{\bibinfo{volume}{D98}},
  \bibinfo{pages}{023020} (\bibinfo{year}{2018}), \eprint{1804.09660}.

\bibitem[{\citenamefont{Dasgupta and Borah}(2014)}]{Dasgupta:2014hha}
\bibinfo{author}{\bibfnamefont{A.}~\bibnamefont{Dasgupta}} \bibnamefont{and}
  \bibinfo{author}{\bibfnamefont{D.}~\bibnamefont{Borah}},
  \bibinfo{journal}{Nucl. Phys.} \textbf{\bibinfo{volume}{B889}},
  \bibinfo{pages}{637} (\bibinfo{year}{2014}), \eprint{1404.5261}.

\bibitem[{\citenamefont{Das et~al.}(2017)\citenamefont{Das, Nomura, Okada, and
  Roy}}]{Das:2017ski}
\bibinfo{author}{\bibfnamefont{A.}~\bibnamefont{Das}},
  \bibinfo{author}{\bibfnamefont{T.}~\bibnamefont{Nomura}},
  \bibinfo{author}{\bibfnamefont{H.}~\bibnamefont{Okada}}, \bibnamefont{and}
  \bibinfo{author}{\bibfnamefont{S.}~\bibnamefont{Roy}},
  \bibinfo{journal}{Phys. Rev.} \textbf{\bibinfo{volume}{D96}},
  \bibinfo{pages}{075001} (\bibinfo{year}{2017}), \eprint{1704.02078}.

\bibitem[{\citenamefont{Deshpande and Ma}(1978)}]{Deshpande:1977rw}
\bibinfo{author}{\bibfnamefont{N.~G.} \bibnamefont{Deshpande}}
  \bibnamefont{and} \bibinfo{author}{\bibfnamefont{E.}~\bibnamefont{Ma}},
  \bibinfo{journal}{Phys. Rev.} \textbf{\bibinfo{volume}{D18}},
  \bibinfo{pages}{2574} (\bibinfo{year}{1978}).

\bibitem[{\citenamefont{Cirelli et~al.}(2006)\citenamefont{Cirelli, Fornengo,
  and Strumia}}]{Cirelli:2005uq}
\bibinfo{author}{\bibfnamefont{M.}~\bibnamefont{Cirelli}},
  \bibinfo{author}{\bibfnamefont{N.}~\bibnamefont{Fornengo}}, \bibnamefont{and}
  \bibinfo{author}{\bibfnamefont{A.}~\bibnamefont{Strumia}},
  \bibinfo{journal}{Nucl. Phys.} \textbf{\bibinfo{volume}{B753}},
  \bibinfo{pages}{178} (\bibinfo{year}{2006}), \eprint{hep-ph/0512090}.

\bibitem[{\citenamefont{Barbieri et~al.}(2006)\citenamefont{Barbieri, Hall, and
  Rychkov}}]{Barbieri:2006dq}
\bibinfo{author}{\bibfnamefont{R.}~\bibnamefont{Barbieri}},
  \bibinfo{author}{\bibfnamefont{L.~J.} \bibnamefont{Hall}}, \bibnamefont{and}
  \bibinfo{author}{\bibfnamefont{V.~S.} \bibnamefont{Rychkov}},
  \bibinfo{journal}{Phys. Rev.} \textbf{\bibinfo{volume}{D74}},
  \bibinfo{pages}{015007} (\bibinfo{year}{2006}), \eprint{hep-ph/0603188}.

\bibitem[{\citenamefont{Ma}(2006{\natexlab{b}})}]{Ma:2006fn}
\bibinfo{author}{\bibfnamefont{E.}~\bibnamefont{Ma}}, \bibinfo{journal}{Mod.
  Phys. Lett.} \textbf{\bibinfo{volume}{A21}}, \bibinfo{pages}{1777}
  (\bibinfo{year}{2006}{\natexlab{b}}), \eprint{hep-ph/0605180}.

\bibitem[{\citenamefont{Lopez~Honorez et~al.}(2007)\citenamefont{Lopez~Honorez,
  Nezri, Oliver, and Tytgat}}]{LopezHonorez:2006gr}
\bibinfo{author}{\bibfnamefont{L.}~\bibnamefont{Lopez~Honorez}},
  \bibinfo{author}{\bibfnamefont{E.}~\bibnamefont{Nezri}},
  \bibinfo{author}{\bibfnamefont{J.~F.} \bibnamefont{Oliver}},
  \bibnamefont{and} \bibinfo{author}{\bibfnamefont{M.~H.~G.}
  \bibnamefont{Tytgat}}, \bibinfo{journal}{JCAP}
  \textbf{\bibinfo{volume}{0702}}, \bibinfo{pages}{028} (\bibinfo{year}{2007}),
  \eprint{hep-ph/0612275}.

\bibitem[{\citenamefont{Hambye et~al.}(2009)\citenamefont{Hambye, Ling,
  Lopez~Honorez, and Rocher}}]{Hambye:2009pw}
\bibinfo{author}{\bibfnamefont{T.}~\bibnamefont{Hambye}},
  \bibinfo{author}{\bibfnamefont{F.~S.} \bibnamefont{Ling}},
  \bibinfo{author}{\bibfnamefont{L.}~\bibnamefont{Lopez~Honorez}},
  \bibnamefont{and} \bibinfo{author}{\bibfnamefont{J.}~\bibnamefont{Rocher}},
  \bibinfo{journal}{JHEP} \textbf{\bibinfo{volume}{07}}, \bibinfo{pages}{090}
  (\bibinfo{year}{2009}), \bibinfo{note}{[Erratum: JHEP05,066(2010)]},
  \eprint{0903.4010}.

\bibitem[{\citenamefont{Dolle and Su}(2009)}]{Dolle:2009fn}
\bibinfo{author}{\bibfnamefont{E.~M.} \bibnamefont{Dolle}} \bibnamefont{and}
  \bibinfo{author}{\bibfnamefont{S.}~\bibnamefont{Su}}, \bibinfo{journal}{Phys.
  Rev.} \textbf{\bibinfo{volume}{D80}}, \bibinfo{pages}{055012}
  (\bibinfo{year}{2009}), \eprint{0906.1609}.

\bibitem[{\citenamefont{Lopez~Honorez and Yaguna}(2010)}]{Honorez:2010re}
\bibinfo{author}{\bibfnamefont{L.}~\bibnamefont{Lopez~Honorez}}
  \bibnamefont{and} \bibinfo{author}{\bibfnamefont{C.~E.}
  \bibnamefont{Yaguna}}, \bibinfo{journal}{JHEP} \textbf{\bibinfo{volume}{09}},
  \bibinfo{pages}{046} (\bibinfo{year}{2010}), \eprint{1003.3125}.

\bibitem[{\citenamefont{Lopez~Honorez and Yaguna}(2011)}]{LopezHonorez:2010tb}
\bibinfo{author}{\bibfnamefont{L.}~\bibnamefont{Lopez~Honorez}}
  \bibnamefont{and} \bibinfo{author}{\bibfnamefont{C.~E.}
  \bibnamefont{Yaguna}}, \bibinfo{journal}{JCAP}
  \textbf{\bibinfo{volume}{1101}}, \bibinfo{pages}{002} (\bibinfo{year}{2011}),
  \eprint{1011.1411}.

\bibitem[{\citenamefont{Gustafsson et~al.}(2012)\citenamefont{Gustafsson,
  Rydbeck, Lopez-Honorez, and Lundstrom}}]{Gustafsson:2012aj}
\bibinfo{author}{\bibfnamefont{M.}~\bibnamefont{Gustafsson}},
  \bibinfo{author}{\bibfnamefont{S.}~\bibnamefont{Rydbeck}},
  \bibinfo{author}{\bibfnamefont{L.}~\bibnamefont{Lopez-Honorez}},
  \bibnamefont{and}
  \bibinfo{author}{\bibfnamefont{E.}~\bibnamefont{Lundstrom}},
  \bibinfo{journal}{Phys. Rev.} \textbf{\bibinfo{volume}{D86}},
  \bibinfo{pages}{075019} (\bibinfo{year}{2012}), \eprint{1206.6316}.

\bibitem[{\citenamefont{Goudelis et~al.}(2013)\citenamefont{Goudelis, Herrmann,
  and Stal}}]{Goudelis:2013uca}
\bibinfo{author}{\bibfnamefont{A.}~\bibnamefont{Goudelis}},
  \bibinfo{author}{\bibfnamefont{B.}~\bibnamefont{Herrmann}}, \bibnamefont{and}
  \bibinfo{author}{\bibfnamefont{O.}~\bibnamefont{Stal}},
  \bibinfo{journal}{JHEP} \textbf{\bibinfo{volume}{09}}, \bibinfo{pages}{106}
  (\bibinfo{year}{2013}), \eprint{1303.3010}.

\bibitem[{\citenamefont{Arhrib et~al.}(2014)\citenamefont{Arhrib, Tsai, Yuan,
  and Yuan}}]{Arhrib:2013ela}
\bibinfo{author}{\bibfnamefont{A.}~\bibnamefont{Arhrib}},
  \bibinfo{author}{\bibfnamefont{Y.-L.~S.} \bibnamefont{Tsai}},
  \bibinfo{author}{\bibfnamefont{Q.}~\bibnamefont{Yuan}}, \bibnamefont{and}
  \bibinfo{author}{\bibfnamefont{T.-C.} \bibnamefont{Yuan}},
  \bibinfo{journal}{JCAP} \textbf{\bibinfo{volume}{1406}}, \bibinfo{pages}{030}
  (\bibinfo{year}{2014}), \eprint{1310.0358}.

\bibitem[{\citenamefont{Díaz et~al.}(2016)\citenamefont{Díaz, Koch, and
  Urrutia-Quiroga}}]{Diaz:2015pyv}
\bibinfo{author}{\bibfnamefont{M.~A.} \bibnamefont{Díaz}},
  \bibinfo{author}{\bibfnamefont{B.}~\bibnamefont{Koch}}, \bibnamefont{and}
  \bibinfo{author}{\bibfnamefont{S.}~\bibnamefont{Urrutia-Quiroga}},
  \bibinfo{journal}{Adv. High Energy Phys.} \textbf{\bibinfo{volume}{2016}},
  \bibinfo{pages}{8278375} (\bibinfo{year}{2016}), \eprint{1511.04429}.

\bibitem[{\citenamefont{Ahriche et~al.}(2018)\citenamefont{Ahriche, Jueid, and
  Nasri}}]{Ahriche:2017iar}
\bibinfo{author}{\bibfnamefont{A.}~\bibnamefont{Ahriche}},
  \bibinfo{author}{\bibfnamefont{A.}~\bibnamefont{Jueid}}, \bibnamefont{and}
  \bibinfo{author}{\bibfnamefont{S.}~\bibnamefont{Nasri}},
  \bibinfo{journal}{Phys. Rev.} \textbf{\bibinfo{volume}{D97}},
  \bibinfo{pages}{095012} (\bibinfo{year}{2018}), \eprint{1710.03824}.

\bibitem[{\citenamefont{Merle and Platscher}(2015)}]{Merle:2015ica}
\bibinfo{author}{\bibfnamefont{A.}~\bibnamefont{Merle}} \bibnamefont{and}
  \bibinfo{author}{\bibfnamefont{M.}~\bibnamefont{Platscher}},
  \bibinfo{journal}{JHEP} \textbf{\bibinfo{volume}{11}}, \bibinfo{pages}{148}
  (\bibinfo{year}{2015}), \eprint{1507.06314}.

\bibitem[{\citenamefont{'t~Hooft}(1980)}]{tHooft:1979rat}
\bibinfo{author}{\bibfnamefont{G.}~\bibnamefont{'t~Hooft}},
  \bibinfo{journal}{NATO Sci. Ser. B} \textbf{\bibinfo{volume}{59}},
  \bibinfo{pages}{135} (\bibinfo{year}{1980}).

\bibitem[{\citenamefont{Esteban et~al.}(2017)\citenamefont{Esteban,
  Gonzalez-Garcia, Maltoni, Martinez-Soler, and Schwetz}}]{Esteban:2016qun}
\bibinfo{author}{\bibfnamefont{I.}~\bibnamefont{Esteban}},
  \bibinfo{author}{\bibfnamefont{M.~C.} \bibnamefont{Gonzalez-Garcia}},
  \bibinfo{author}{\bibfnamefont{M.}~\bibnamefont{Maltoni}},
  \bibinfo{author}{\bibfnamefont{I.}~\bibnamefont{Martinez-Soler}},
  \bibnamefont{and} \bibinfo{author}{\bibfnamefont{T.}~\bibnamefont{Schwetz}},
  \bibinfo{journal}{JHEP} \textbf{\bibinfo{volume}{01}}, \bibinfo{pages}{087}
  (\bibinfo{year}{2017}), \eprint{1611.01514}.

\bibitem[{\citenamefont{Casas and Ibarra}(2001)}]{Casas:2001sr}
\bibinfo{author}{\bibfnamefont{J.~A.} \bibnamefont{Casas}} \bibnamefont{and}
  \bibinfo{author}{\bibfnamefont{A.}~\bibnamefont{Ibarra}},
  \bibinfo{journal}{Nucl. Phys.} \textbf{\bibinfo{volume}{B618}},
  \bibinfo{pages}{171} (\bibinfo{year}{2001}), \eprint{hep-ph/0103065}.

\bibitem[{\citenamefont{Birrell and Davies}(1984)}]{Birrell:1982ix}
\bibinfo{author}{\bibfnamefont{N.~D.} \bibnamefont{Birrell}} \bibnamefont{and}
  \bibinfo{author}{\bibfnamefont{P.~C.~W.} \bibnamefont{Davies}},
  \emph{\bibinfo{title}{{Quantum Fields in Curved Space}}}, Cambridge
  Monographs on Mathematical Physics (\bibinfo{publisher}{Cambridge Univ.
  Press}, \bibinfo{address}{Cambridge, UK}, \bibinfo{year}{1984}).

\bibitem[{\citenamefont{Capozziello et~al.}(1997)\citenamefont{Capozziello,
  de~Ritis, and Marino}}]{Capozziello:1996xg}
\bibinfo{author}{\bibfnamefont{S.}~\bibnamefont{Capozziello}},
  \bibinfo{author}{\bibfnamefont{R.}~\bibnamefont{de~Ritis}}, \bibnamefont{and}
  \bibinfo{author}{\bibfnamefont{A.~A.} \bibnamefont{Marino}},
  \bibinfo{journal}{Class. Quant. Grav.} \textbf{\bibinfo{volume}{14}},
  \bibinfo{pages}{3243} (\bibinfo{year}{1997}), \eprint{gr-qc/9612053}.

\bibitem[{\citenamefont{Kaiser}(2010)}]{Kaiser:2010ps}
\bibinfo{author}{\bibfnamefont{D.~I.} \bibnamefont{Kaiser}},
  \bibinfo{journal}{Phys. Rev.} \textbf{\bibinfo{volume}{D81}},
  \bibinfo{pages}{084044} (\bibinfo{year}{2010}), \eprint{1003.1159}.

\bibitem[{\citenamefont{Starobinsky}(1979)}]{Starobinsky:1979ty}
\bibinfo{author}{\bibfnamefont{A.~A.} \bibnamefont{Starobinsky}},
  \bibinfo{journal}{JETP Lett.} \textbf{\bibinfo{volume}{30}},
  \bibinfo{pages}{682} (\bibinfo{year}{1979}), \bibinfo{note}{[,767(1979)]}.

\bibitem[{\citenamefont{Calmet and Kuntz}(2016)}]{Calmet:2016fsr}
\bibinfo{author}{\bibfnamefont{X.}~\bibnamefont{Calmet}} \bibnamefont{and}
  \bibinfo{author}{\bibfnamefont{I.}~\bibnamefont{Kuntz}},
  \bibinfo{journal}{Eur. Phys. J.} \textbf{\bibinfo{volume}{C76}},
  \bibinfo{pages}{289} (\bibinfo{year}{2016}), \eprint{1605.02236}.

\bibitem[{\citenamefont{Remmen and Carroll}(2014)}]{Remmen:2014mia}
\bibinfo{author}{\bibfnamefont{G.~N.} \bibnamefont{Remmen}} \bibnamefont{and}
  \bibinfo{author}{\bibfnamefont{S.~M.} \bibnamefont{Carroll}},
  \bibinfo{journal}{Phys. Rev.} \textbf{\bibinfo{volume}{D90}},
  \bibinfo{pages}{063517} (\bibinfo{year}{2014}), \eprint{1405.5538}.

\bibitem[{\citenamefont{Liddle and Leach}(2003)}]{Liddle:2003as}
\bibinfo{author}{\bibfnamefont{A.~R.} \bibnamefont{Liddle}} \bibnamefont{and}
  \bibinfo{author}{\bibfnamefont{S.~M.} \bibnamefont{Leach}},
  \bibinfo{journal}{Phys. Rev.} \textbf{\bibinfo{volume}{D68}},
  \bibinfo{pages}{103503} (\bibinfo{year}{2003}), \eprint{astro-ph/0305263}.

\bibitem[{\citenamefont{Allahverdi et~al.}(2010)\citenamefont{Allahverdi,
  Brandenberger, Cyr-Racine, and Mazumdar}}]{Allahverdi:2010xz}
\bibinfo{author}{\bibfnamefont{R.}~\bibnamefont{Allahverdi}},
  \bibinfo{author}{\bibfnamefont{R.}~\bibnamefont{Brandenberger}},
  \bibinfo{author}{\bibfnamefont{F.-Y.} \bibnamefont{Cyr-Racine}},
  \bibnamefont{and} \bibinfo{author}{\bibfnamefont{A.}~\bibnamefont{Mazumdar}},
  \bibinfo{journal}{Ann. Rev. Nucl. Part. Sci.} \textbf{\bibinfo{volume}{60}},
  \bibinfo{pages}{27} (\bibinfo{year}{2010}), \eprint{1001.2600}.

\bibitem[{\citenamefont{Bezrukov et~al.}(2009)\citenamefont{Bezrukov, Gorbunov,
  and Shaposhnikov}}]{Bezrukov:2008ut}
\bibinfo{author}{\bibfnamefont{F.}~\bibnamefont{Bezrukov}},
  \bibinfo{author}{\bibfnamefont{D.}~\bibnamefont{Gorbunov}}, \bibnamefont{and}
  \bibinfo{author}{\bibfnamefont{M.}~\bibnamefont{Shaposhnikov}},
  \bibinfo{journal}{JCAP} \textbf{\bibinfo{volume}{0906}}, \bibinfo{pages}{029}
  (\bibinfo{year}{2009}), \eprint{0812.3622}.

\bibitem[{\citenamefont{Garcia-Bellido
  et~al.}(2009)\citenamefont{Garcia-Bellido, Figueroa, and
  Rubio}}]{GarciaBellido:2008ab}
\bibinfo{author}{\bibfnamefont{J.}~\bibnamefont{Garcia-Bellido}},
  \bibinfo{author}{\bibfnamefont{D.~G.} \bibnamefont{Figueroa}},
  \bibnamefont{and} \bibinfo{author}{\bibfnamefont{J.}~\bibnamefont{Rubio}},
  \bibinfo{journal}{Phys. Rev.} \textbf{\bibinfo{volume}{D79}},
  \bibinfo{pages}{063531} (\bibinfo{year}{2009}), \eprint{0812.4624}.

\bibitem[{\citenamefont{Repond and Rubio}(2016)}]{Repond:2016sol}
\bibinfo{author}{\bibfnamefont{J.}~\bibnamefont{Repond}} \bibnamefont{and}
  \bibinfo{author}{\bibfnamefont{J.}~\bibnamefont{Rubio}},
  \bibinfo{journal}{JCAP} \textbf{\bibinfo{volume}{1607}}, \bibinfo{pages}{043}
  (\bibinfo{year}{2016}), \eprint{1604.08238}.

\bibitem[{\citenamefont{Gondolo and Gelmini}(1991)}]{Gondolo:1990dk}
\bibinfo{author}{\bibfnamefont{P.}~\bibnamefont{Gondolo}} \bibnamefont{and}
  \bibinfo{author}{\bibfnamefont{G.}~\bibnamefont{Gelmini}},
  \bibinfo{journal}{Nucl. Phys.} \textbf{\bibinfo{volume}{B360}},
  \bibinfo{pages}{145} (\bibinfo{year}{1991}).

\bibitem[{\citenamefont{Griest and Seckel}(1991)}]{Griest:1990kh}
\bibinfo{author}{\bibfnamefont{K.}~\bibnamefont{Griest}} \bibnamefont{and}
  \bibinfo{author}{\bibfnamefont{D.}~\bibnamefont{Seckel}},
  \bibinfo{journal}{Phys. Rev.} \textbf{\bibinfo{volume}{D43}},
  \bibinfo{pages}{3191} (\bibinfo{year}{1991}).

\bibitem[{\citenamefont{Borah et~al.}(2017)\citenamefont{Borah, Sadhukhan, and
  Sahoo}}]{Borah:2017dqx}
\bibinfo{author}{\bibfnamefont{D.}~\bibnamefont{Borah}},
  \bibinfo{author}{\bibfnamefont{S.}~\bibnamefont{Sadhukhan}},
  \bibnamefont{and} \bibinfo{author}{\bibfnamefont{S.}~\bibnamefont{Sahoo}}
  (\bibinfo{year}{2017}), \eprint{1703.08674}.

\bibitem[{\citenamefont{Akerib et~al.}(2015)}]{Akerib:2015cja}
\bibinfo{author}{\bibfnamefont{D.~S.} \bibnamefont{Akerib}}
  \bibnamefont{et~al.} (\bibinfo{collaboration}{LZ}) (\bibinfo{year}{2015}),
  \eprint{1509.02910}.

\bibitem[{\citenamefont{Aprile et~al.}(2016)}]{Aprile:2015uzo}
\bibinfo{author}{\bibfnamefont{E.}~\bibnamefont{Aprile}} \bibnamefont{et~al.}
  (\bibinfo{collaboration}{XENON}), \bibinfo{journal}{JCAP}
  \textbf{\bibinfo{volume}{1604}}, \bibinfo{pages}{027} (\bibinfo{year}{2016}),
  \eprint{1512.07501}.

\bibitem[{\citenamefont{Aalbers et~al.}(2016)}]{Aalbers:2016jon}
\bibinfo{author}{\bibfnamefont{J.}~\bibnamefont{Aalbers}} \bibnamefont{et~al.}
  (\bibinfo{collaboration}{DARWIN}), \bibinfo{journal}{JCAP}
  \textbf{\bibinfo{volume}{1611}}, \bibinfo{pages}{017} (\bibinfo{year}{2016}),
  \eprint{1606.07001}.

\bibitem[{\citenamefont{Liu et~al.}(2017)\citenamefont{Liu, Chen, and
  Ji}}]{Liu:2017drf}
\bibinfo{author}{\bibfnamefont{J.}~\bibnamefont{Liu}},
  \bibinfo{author}{\bibfnamefont{X.}~\bibnamefont{Chen}}, \bibnamefont{and}
  \bibinfo{author}{\bibfnamefont{X.}~\bibnamefont{Ji}},
  \bibinfo{journal}{Nature Phys.} \textbf{\bibinfo{volume}{13}},
  \bibinfo{pages}{212} (\bibinfo{year}{2017}), \eprint{1709.00688}.

\bibitem[{\citenamefont{Kashiwase and Suematsu}(2012)}]{Kashiwase:2012xd}
\bibinfo{author}{\bibfnamefont{S.}~\bibnamefont{Kashiwase}} \bibnamefont{and}
  \bibinfo{author}{\bibfnamefont{D.}~\bibnamefont{Suematsu}},
  \bibinfo{journal}{Phys. Rev.} \textbf{\bibinfo{volume}{D86}},
  \bibinfo{pages}{053001} (\bibinfo{year}{2012}), \eprint{1207.2594}.

\bibitem[{\citenamefont{Kashiwase and Suematsu}(2013)}]{Kashiwase:2013uy}
\bibinfo{author}{\bibfnamefont{S.}~\bibnamefont{Kashiwase}} \bibnamefont{and}
  \bibinfo{author}{\bibfnamefont{D.}~\bibnamefont{Suematsu}},
  \bibinfo{journal}{Eur. Phys. J.} \textbf{\bibinfo{volume}{C73}},
  \bibinfo{pages}{2484} (\bibinfo{year}{2013}), \eprint{1301.2087}.

\bibitem[{\citenamefont{Racker}(2014)}]{Racker:2013lua}
\bibinfo{author}{\bibfnamefont{J.}~\bibnamefont{Racker}},
  \bibinfo{journal}{JCAP} \textbf{\bibinfo{volume}{1403}}, \bibinfo{pages}{025}
  (\bibinfo{year}{2014}), \eprint{1308.1840}.

\bibitem[{\citenamefont{Clarke et~al.}(2015)\citenamefont{Clarke, Foot, and
  Volkas}}]{Clarke:2015hta}
\bibinfo{author}{\bibfnamefont{J.~D.} \bibnamefont{Clarke}},
  \bibinfo{author}{\bibfnamefont{R.}~\bibnamefont{Foot}}, \bibnamefont{and}
  \bibinfo{author}{\bibfnamefont{R.~R.} \bibnamefont{Volkas}},
  \bibinfo{journal}{Phys. Rev.} \textbf{\bibinfo{volume}{D92}},
  \bibinfo{pages}{033006} (\bibinfo{year}{2015}), \eprint{1505.05744}.

\bibitem[{\citenamefont{Buchmuller et~al.}(2005)\citenamefont{Buchmuller,
  Di~Bari, and Plumacher}}]{Buchmuller:2004nz}
\bibinfo{author}{\bibfnamefont{W.}~\bibnamefont{Buchmuller}},
  \bibinfo{author}{\bibfnamefont{P.}~\bibnamefont{Di~Bari}}, \bibnamefont{and}
  \bibinfo{author}{\bibfnamefont{M.}~\bibnamefont{Plumacher}},
  \bibinfo{journal}{Annals Phys.} \textbf{\bibinfo{volume}{315}},
  \bibinfo{pages}{305} (\bibinfo{year}{2005}), \eprint{hep-ph/0401240}.

\bibitem[{\citenamefont{Davidson and Ibarra}(2002)}]{Davidson:2002qv}
\bibinfo{author}{\bibfnamefont{S.}~\bibnamefont{Davidson}} \bibnamefont{and}
  \bibinfo{author}{\bibfnamefont{A.}~\bibnamefont{Ibarra}},
  \bibinfo{journal}{Phys. Lett.} \textbf{\bibinfo{volume}{B535}},
  \bibinfo{pages}{25} (\bibinfo{year}{2002}), \eprint{hep-ph/0202239}.

\bibitem[{\citenamefont{Buchmuller et~al.}(2002)\citenamefont{Buchmuller,
  Di~Bari, and Plumacher}}]{Buchmuller:2002rq}
\bibinfo{author}{\bibfnamefont{W.}~\bibnamefont{Buchmuller}},
  \bibinfo{author}{\bibfnamefont{P.}~\bibnamefont{Di~Bari}}, \bibnamefont{and}
  \bibinfo{author}{\bibfnamefont{M.}~\bibnamefont{Plumacher}},
  \bibinfo{journal}{Nucl. Phys.} \textbf{\bibinfo{volume}{B643}},
  \bibinfo{pages}{367} (\bibinfo{year}{2002}), \bibinfo{note}{[Erratum: Nucl.
  Phys.B793,362(2008)]}, \eprint{hep-ph/0205349}.

\bibitem[{\citenamefont{Moffat et~al.}(2018)\citenamefont{Moffat, Pascoli,
  Petcov, Schulz, and Turner}}]{Moffat:2018wke}
\bibinfo{author}{\bibfnamefont{K.}~\bibnamefont{Moffat}},
  \bibinfo{author}{\bibfnamefont{S.}~\bibnamefont{Pascoli}},
  \bibinfo{author}{\bibfnamefont{S.~T.} \bibnamefont{Petcov}},
  \bibinfo{author}{\bibfnamefont{H.}~\bibnamefont{Schulz}}, \bibnamefont{and}
  \bibinfo{author}{\bibfnamefont{J.}~\bibnamefont{Turner}},
  \bibinfo{journal}{Phys. Rev.} \textbf{\bibinfo{volume}{D98}},
  \bibinfo{pages}{015036} (\bibinfo{year}{2018}), \eprint{1804.05066}.

\bibitem[{\citenamefont{Pilaftsis and Underwood}(2004)}]{Pilaftsis:2003gt}
\bibinfo{author}{\bibfnamefont{A.}~\bibnamefont{Pilaftsis}} \bibnamefont{and}
  \bibinfo{author}{\bibfnamefont{T.~E.~J.} \bibnamefont{Underwood}},
  \bibinfo{journal}{Nucl. Phys.} \textbf{\bibinfo{volume}{B692}},
  \bibinfo{pages}{303} (\bibinfo{year}{2004}), \eprint{hep-ph/0309342}.

\bibitem[{\citenamefont{Dev et~al.}(2018)\citenamefont{Dev, Garny, Klaric,
  Millington, and Teresi}}]{Dev:2017wwc}
\bibinfo{author}{\bibfnamefont{P.~S.~B.} \bibnamefont{Dev}},
  \bibinfo{author}{\bibfnamefont{M.}~\bibnamefont{Garny}},
  \bibinfo{author}{\bibfnamefont{J.}~\bibnamefont{Klaric}},
  \bibinfo{author}{\bibfnamefont{P.}~\bibnamefont{Millington}},
  \bibnamefont{and} \bibinfo{author}{\bibfnamefont{D.}~\bibnamefont{Teresi}},
  \bibinfo{journal}{Int. J. Mod. Phys.} \textbf{\bibinfo{volume}{A33}},
  \bibinfo{pages}{1842003} (\bibinfo{year}{2018}), \eprint{1711.02863}.

\bibitem[{\citenamefont{Branco et~al.}(2012)\citenamefont{Branco, Ferreira,
  Lavoura, Rebelo, Sher, and Silva}}]{Branco:2011iw}
\bibinfo{author}{\bibfnamefont{G.~C.} \bibnamefont{Branco}},
  \bibinfo{author}{\bibfnamefont{P.~M.} \bibnamefont{Ferreira}},
  \bibinfo{author}{\bibfnamefont{L.}~\bibnamefont{Lavoura}},
  \bibinfo{author}{\bibfnamefont{M.~N.} \bibnamefont{Rebelo}},
  \bibinfo{author}{\bibfnamefont{M.}~\bibnamefont{Sher}}, \bibnamefont{and}
  \bibinfo{author}{\bibfnamefont{J.~P.} \bibnamefont{Silva}},
  \bibinfo{journal}{Phys. Rept.} \textbf{\bibinfo{volume}{516}},
  \bibinfo{pages}{1} (\bibinfo{year}{2012}), \eprint{1106.0034}.

\bibitem[{\citenamefont{Chakrabarty et~al.}(2015)\citenamefont{Chakrabarty,
  Ghosh, Mukhopadhyaya, and Saha}}]{Chakrabarty:2015yia}
\bibinfo{author}{\bibfnamefont{N.}~\bibnamefont{Chakrabarty}},
  \bibinfo{author}{\bibfnamefont{D.~K.} \bibnamefont{Ghosh}},
  \bibinfo{author}{\bibfnamefont{B.}~\bibnamefont{Mukhopadhyaya}},
  \bibnamefont{and} \bibinfo{author}{\bibfnamefont{I.}~\bibnamefont{Saha}},
  \bibinfo{journal}{Phys. Rev.} \textbf{\bibinfo{volume}{D92}},
  \bibinfo{pages}{015002} (\bibinfo{year}{2015}), \eprint{1501.03700}.

\bibitem[{\citenamefont{Staub}(2008)}]{Staub:2008uz}
\bibinfo{author}{\bibfnamefont{F.}~\bibnamefont{Staub}} (\bibinfo{year}{2008}),
  \eprint{0806.0538}.

\bibitem[{\citenamefont{Ade et~al.}(2014)}]{Ade:2014xna}
\bibinfo{author}{\bibfnamefont{P.~A.~R.} \bibnamefont{Ade}}
  \bibnamefont{et~al.} (\bibinfo{collaboration}{BICEP2}),
  \bibinfo{journal}{Phys. Rev. Lett.} \textbf{\bibinfo{volume}{112}},
  \bibinfo{pages}{241101} (\bibinfo{year}{2014}), \eprint{1403.3985}.

\bibitem[{\citenamefont{Ade et~al.}(2015)}]{Ade:2015tva}
\bibinfo{author}{\bibfnamefont{P.~A.~R.} \bibnamefont{Ade}}
  \bibnamefont{et~al.} (\bibinfo{collaboration}{BICEP2, Planck}),
  \bibinfo{journal}{Phys. Rev. Lett.} \textbf{\bibinfo{volume}{114}},
  \bibinfo{pages}{101301} (\bibinfo{year}{2015}), \eprint{1502.00612}.

\end{thebibliography}

\end{document}